\documentclass[aps,prl,twocolumn,superscriptaddress,floatfix,amsmath,amssymb,amsfonts, showpacs,preprintnumbers]{revtex4}
\usepackage{hhline}
\usepackage{hyperref}
\usepackage{physics}
\usepackage{float}
\usepackage{graphicx}% Include figure files
\usepackage{dcolumn}% Align table columns on decimal point
\usepackage{bm}% bold math
\usepackage{color}
\newcommand{\ps}{}

\begin{document}

\title{Identifying Majorana bound states by tunneling shot-noise tomography}% Force line breaks with \\

\author{Vivien Perrin }
\affiliation{%
Universit\'e Paris-Saclay, CNRS, Laboratoire de Physique des Solides, 91405, Orsay, France
}%

\author{Marcello Civelli }
\affiliation{%
Universit\'e Paris-Saclay, CNRS, Laboratoire de Physique des Solides, 91405, Orsay, France
}%
\author{Pascal Simon }
\affiliation{%
Universit\'e Paris-Saclay, CNRS, Laboratoire de Physique des Solides, 91405, Orsay, France
}%

%\date{\today}% It is always \today, today,
             %  but any date may be explicitly specified

\begin{abstract}
%  Majorana fermions are  promising building blocks of forthcoming
%    technology in quantum computing.
    %However, %in spite of the huge efforts from the wide condensed matter community,
%    Their proper identification has remained however a difficult issue, because of concomitant competition with other
%    topologically trivial fermionic states, which can poison the Majorana fermion detection in most spectroscopic probes.
	Majorana fermions are  promising building blocks of forthcoming
  technology in quantum computing.
	However, their non-ambiguous identification 
	has remained  a difficult issue because of the concomitant competition with other
    topologically trivial fermionic states, which poison their detection in most spectroscopic probes.
		By employing numerical and analytical methods, here we show that the Fano factor tomography
    is a key distinctive feature of a Majorana bound state,
      displaying a spatially constant Poissonian value equal to one.
      In contrast, the Fano factor of other trivial fermionic states,
      like the Yu-Shiba-Rusinov or Andreev ones, is strongly spatially dependent and exceeds one as 
		 a direct consequence of the local particle-hole symmetry breaking.
\end{abstract}

\maketitle

%\tableofcontents

%
\begin{figure}[t]
    \centering
    \includegraphics[width=8cm]{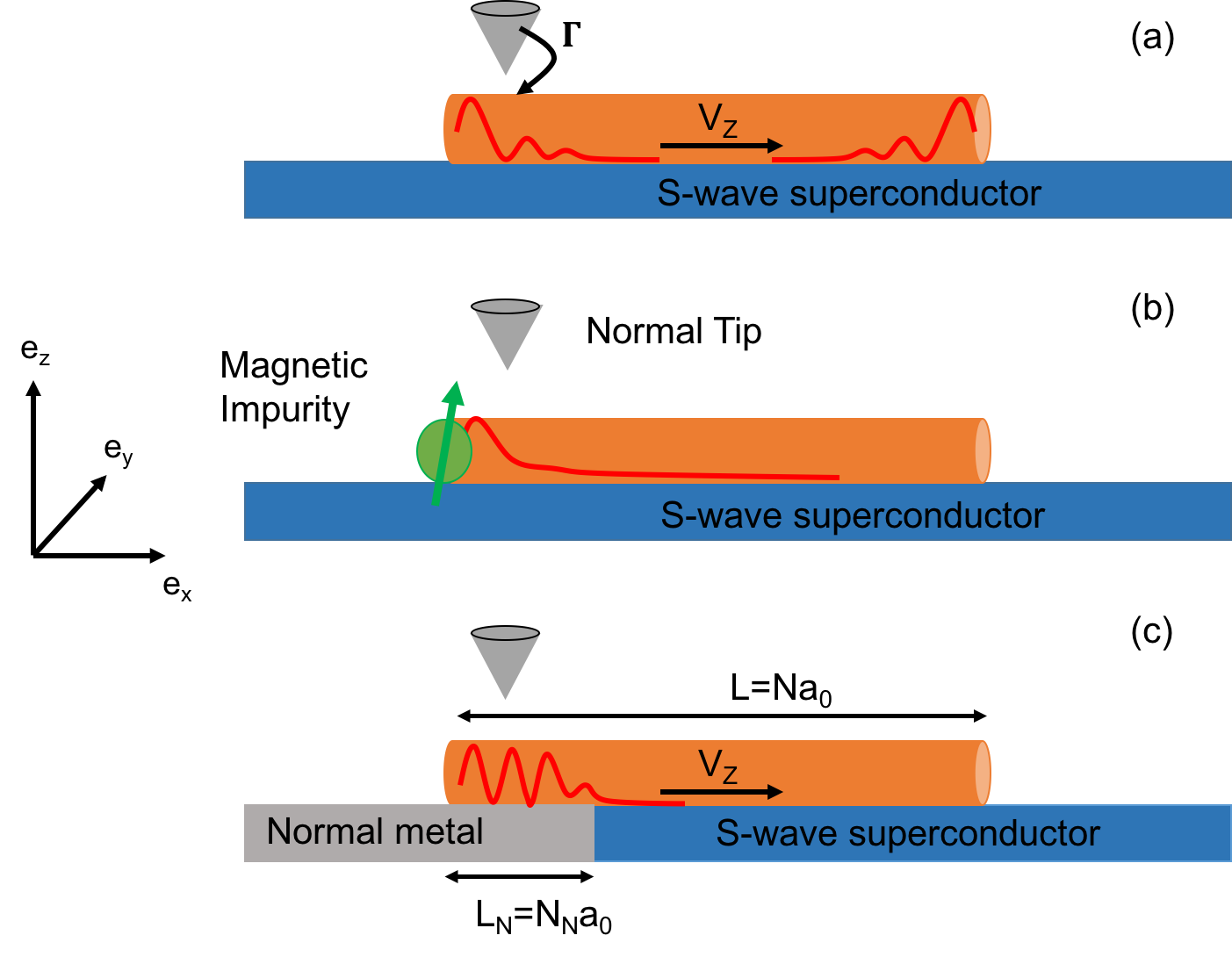}
    \caption{Schematic representation of the typical experimental set-up considered. (a) A nano-wire  placed on top of a s-wave superconductor subject to an external Zeeman field $V_Z$ along the wire {$\mathbf{e}_x$} axis. This system can host MBS at its ends, as depicted by the red solid lines.
    (b) A nano-wire terminated by a magnetic impurity on top of a s-wave superconductor. This system hosts a YSR bound-state at its left-end (red solid line). (c) Same as in a)  except that the substrate contains a small normal region.  This system can host ABS (red solid lines). In all cases, the wire is probed by a metallic tip via tunneling effect of strength $\Gamma$.}
    \label{Schema}
\end{figure}

Since the theoretical work of Kitaev \cite{Kitaev2000} 
	the search of Majorana bound-states (MBS) has been a highly investigated topic in condensed matter. MBS are promising building blocks for quantum information processing \cite{Alicea2010} because they are protected from local perturbation and possess non-Abelian braiding statistics.
Many platforms exhibiting MBS have been experimentally studied in the past decade. This includes semiconducting wires in proximity of a \textit{s}-wave superconductor \cite{Lutchyn2018}, vortices in iron-based superconductors \cite{Wang2018,Chiu2020}, two-dimensional (2D)   phase-controlled Josephson junctions \cite{Ren2019,Fornieri2019}, and  one-dimensional (1D) chains of magnetic atoms on top  of a superconducting substrate \cite{NP2014,Ruby2015,Pawlak2016,Yazdani2017,Kim2018}, to list a few.

Recent scanning tunneling spectroscopy (STS) on self-assembled magnetic chains or magnetic atoms on a \textit{s}-wave superconducting surface
have revealed the existence of zero bias peaks, spatially localized at the end of such
chains \cite{NP2014,Ruby2015,Pawlak2016,Yazdani2017,Kim2018}. These measurements have been interpreted as signatures of topologically protected MBS.
However this conclusion remains openly debated.
%However, finding unambiguous  experimental signatures of those topological states  remains challenging.
Indeed, zero-bias peaks could also be due to the presence of other trivial zero-energy fermionic bound states mimicking topologically protected MBS \cite{Ruby2015}, like, e.g., Yu-Shiba-Rusinov (YSR) bound states \cite{Balatsky2006}, Andreev bound states (ABS) and quasi-Majorana states (QMS)  \cite{Kells2012,Prada2012,Liu2017,Tewari2018,Wimmer2019,Rossi2020,Prada2020}.

Therefore, having clear experimental protocols capable to distinguish trivial fermionic bound-states from MBS
is highly desirable. 
%Many protocols have been proposed to distinguish topological states from trivial ones. 
In this respect,  conductance tomography has been proposed as a promising route, 
as the zero-bias conductance peak due to a pure MBS is quantized. Such a quantization is however weak against quasiparticle poisoning \cite{ZeroBiasPeak,DasSarma2021} or inhomogeneities of the superconducting order parameter \cite{Trauzettel2018}
and moreover necessitates  a strong tunneling regime
to observe a distinctive conductance saturation-plateau in proximity of MBS \cite{Chevallier}.
  
Following earlier theoretical predictions about non-trivial spin signatures of MBS \cite{Sticlet2012,Oreg2015,Black-Schaffer2015,Kotetes2015,Setiawan2015,Szumniak2017} and spin-selective Andreev reflection \cite{Law2014,Sun2016}, Jeon {\it et al.} have used spin polarized
STS 
as a diagnostic tool 
\cite{Jeon2017}. However, discerning some local excess of polarization over a magnetic background
remains a difficult task. 
Other possible signatures of MBS could be provided by current shot-noise~\cite{KTLaw}, spin-resolved shot-noise 
measurements \cite{Oreg2015,Devillard}, finite-frequency current shot-noise~\cite{T.Martin}, or time-resolved transport spectroscopy~\cite{Tuovinen2019}. However, such protocols 
are more involved than the ones employed in usual STS.

Here we focus on the most typical MBS experimental set-up, the hybrid magnetic-superconducting wire.
We show that the current shot-noise spatial tomography with a metallic tip,
nowadays realized within scanning tunneling microscopy (STM) \cite{Massee2018,Allan2018},
can distinguish MBS from other zero-energy fermionic states.
In particular, we shall prove that the MBS Fano factor is spatially constant and equals to one,
the Poissonian limit. This result is a consequence of the local particle-hole symmetry of the
MBS wave function. However, in other zero-energy trivial bound states, like YSR or ABS,
the breaking of such a symmetry implies a strongly spatially dependent Fano factor. 
Even for the case of QMS, we shall show that the Fano
factor oscillates above 1 in the superconducting part
of the wire. Although our proposal cannot definitely distinguish
whether MBS are topologically protected \cite{Prada2020}, the spatially unity Fano factor is a
clear distinguishing signature of the Majorana wave function.

%%%%%%%%%%%%%

{\em Models and methods.}
The typical system that we consider consists of a nanowire deposited on a conventional \textit{s}-wave
  superconductor \cite{Lutchyn2010,Oreg2010} [see Fig. 1(a)] in presence of 
 spin orbit interactions (SOI) and a Zeeman field. 
We first study a zero-energy YSR state created by a magnetic impurity
located at one extremity of the wire [as depicted in Fig. 1(b)]. As second reference case, we study  ABS which can be obtained by assuming the $N_N$ first sites of the wire in contact with a normal substrate [Fig. 1(c)]. 
This situation can also harbor a QMS by fine-tuning a confining potential separating the normal and superconducting part.

A tight-binding Bogoliubov-de Gennes (BdG) Hamiltonian describing such configuration reads 

\begin{align}
    \mathcal{H}_{S}=&\frac{1}{2}
    \sum_{l=0}^{N-1}
    \psi^\dagger_{l,S}[(2t-\mu+V(l))\tau_z+\Delta(l)\tau_x+V_Z\sigma_x]\psi_{l,S}\nonumber\\&
		+\frac{1}{2}\sum_{l=0}^{N-2}\psi^\dagger_{l+1,S}[-{t_w}\tau_z-i\alpha\sigma_y\tau_z]\psi_{l,S} + \text{H.c.}\nonumber\\
    &-\frac{1}{2}\psi^\dagger_{0,S} J\sigma_z \psi_{0,S}~,
\end{align}
where $N$ is the total number of chain sites.
Here we define the Nambu-spinor \hbox{$\psi_{l,S}=(c_{\uparrow,l},c_{\downarrow,l},c^\dagger_{\downarrow,l},-c^\dagger_{\uparrow,l})^T$,} where the operator $c_{\sigma,l}$ annihilates an electron of spin $\sigma$ at site $l$ of the nanowire.
$\Delta(l)=\Delta\,\Theta({l+1-N_N})$ represents a non-uniform pairing potential  induced by proximity effect with the superconducting substrate, $\Theta({l+1-N_N})$ is the Heaviside step function  and $N_N$ is the size of the normal region of the wire. $V(l)=V_0\exp(-\frac{(l-N_N)^2}{2\sigma^2})$ is a smooth Gaussian potential barrier at the NS interface. The Pauli matrices $\tau_i$ and $\sigma_i$ act in the particle-hole and spin spaces respectively, $\mu$  denotes the chemical potential, $t_w$ the nearest-neighbor hopping, $V_Z$ the Zeeman exchange energy, $\alpha$ the SOI. Finally, $J$ is the local exchange coupling due to a magnetic atom localized on the left end of the wire (Fig. 1b). This model is rather general, but we shall
focus on 6 experimentally relevant cases \cite{Chevallier,Oreg2015,Wimmer2019}, described in Table I. %%%%%%%%%%%%%%%%%%%%%%%%%%%%%%%%%%%%%%%
For $N_N=0$, the whole wire is superconducting. When  $J=0$ and $V_Z>\sqrt{\mu^2+\Delta_0^2}$, (case {\it a}), a topological phase characterized by MBS at the ends of the wire emerges (Fig. 1a) \cite{Lutchyn2010,Oreg2010}. 
When $J\neq0$, (case {\it b}), a YSR bound state appears on the wire left end [Fig. 1(b)].
For $N_N>0$, the left part of the wire is in normal state. In this case, when $V_Z>\sqrt{\mu^2+\Delta_0^2}$, (case {\it c}), MBS are present  while for $V_Z<\sqrt{\mu^2+\Delta_0^2}$ a trivial zero-energy ABS localized in the normal region can appear due to fine tuning of parameters. Finally, when $V_Z>\sqrt{\mu^2+\Delta_0^2}$ and a smooth potential barrier is present at the NS interface (case {\it e}), MBS are present while for $V_Z<\sqrt{\mu^2+\Delta_0^2}$,  a zero-energy QMS can appear due to the smooth confinement potential (case {\it f})~\cite{Wimmer2019,Prada2020,SuppMat}. Note that this Hamiltonian is also suitable to describe a ferromagnetic nanowire grown on a substrate with  strong SOI. This situation occurs, e.g., in  Pb \cite{NP2014,Ruby2015,Pawlak2016} or in an array of magnetic impurities adsorbed on a superconducting substrate \cite{Kim2018} displaying a helical magnetic ground state \cite{Choy2011,NP2013,Braunecker2013,Klinovaja2013,Vazifeh2013,Pientka2013,Pientka2014,Poyhonen2014,Ojanen2015a,Peng2015,Hui2015,Braunecker2015}. In such a situation $V_Z$ describes the magnetic exchange energy.

\begin{table}
\label{TableParams}
\begin{center}
\begin{tabular}{|c|c|c|c|c|c|c|c|c|c|c|c|c|c|}
%%%%%%%%%%%%%%%%%%%%%%%%%%%%%%%%%%%%
   \hhline{|=============|}
          &  Case & $\mu$ & $t_w$ & $\Delta$ & $V_Z$ & $\alpha$ & $J$ &$N$&$N_N$&$V_0$&$s$&$\Delta_{\rm eff}$\\\hline
 {\it a} &    MBS 1 &0.5 & 10 & 1 & 2 &  1.2  &0& 80&0&0&X&0.63 \\
 {\it b} &    YSR & 0.5 & 10 & 0.6 & 0  & 1.2 & 11.23 & 80&0&0&X&0.6 \\
 {\it c} &    MBS 2 & 0.5 & 10 & 1 & 1.4 & 2  & 0& 100&20&0&X&0.31\\
 {\it d} &    ABS  & 0.5 & 10 & 1 & 0.38 & 2  & 0 & 100&20&0&X&0.27\\
 {\it e} &    MBS 3  & 1 & 10 & 2 & 3.2 & 3  & 0 & 100&20&4.5&5&0.27\\
 {\it f} &    QMS  & 3 & 10 & 1 & 2.75 & 3  & 0 & 100&20&4.5&5&0.27\\
\hhline{|=============|} \end{tabular}
\end{center}
\caption{Table summarizing 6 different parameter sets of experimentally relevant situations. 
   $\Delta_{\rm eff}$ is the effective gap separating the zero-energy bound-state from other states.} 
\end{table}

\begin{figure*}[t]
    \includegraphics[width=0.85\textwidth]{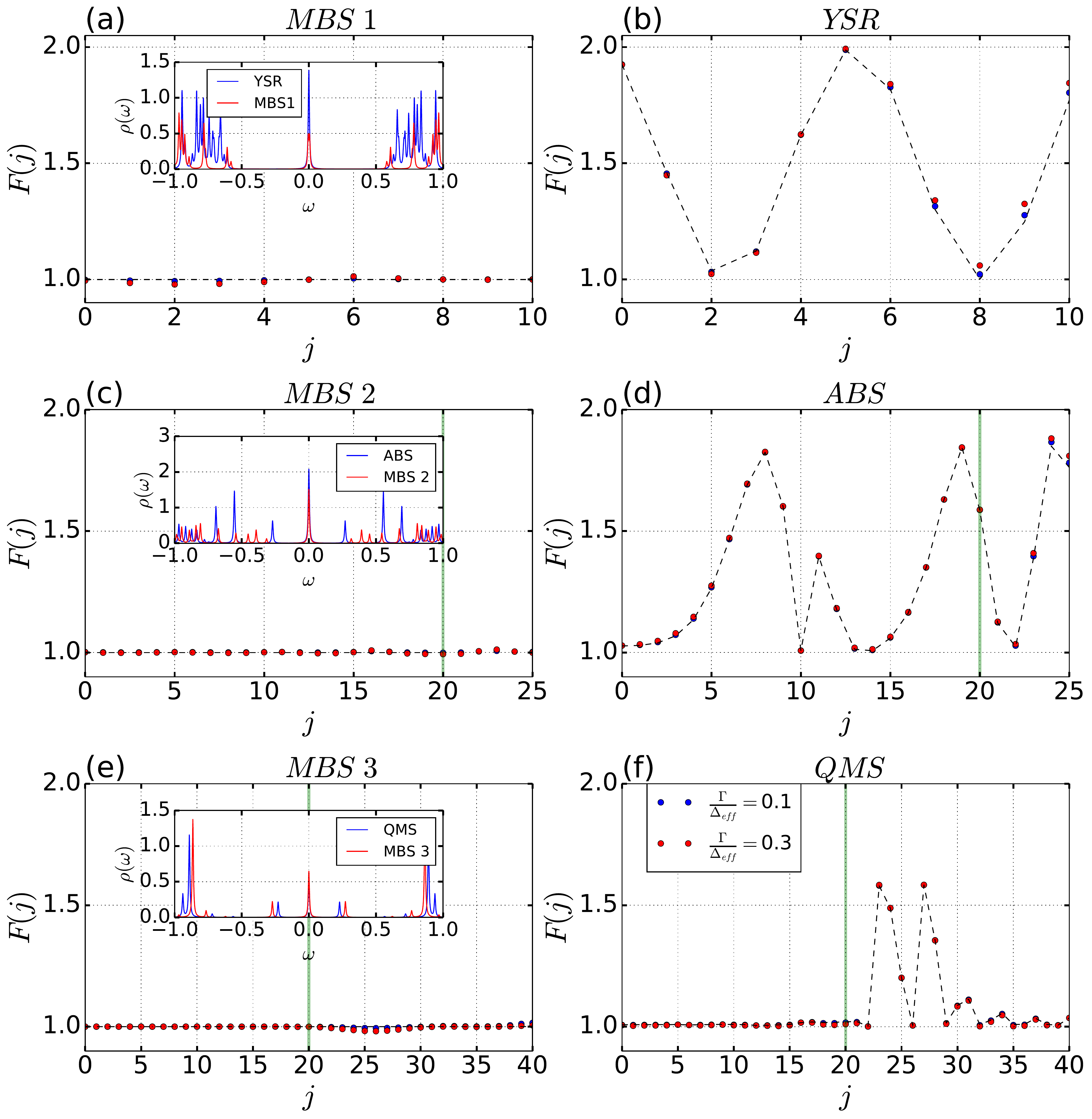}  
    \caption{Dots: Fano factor  as a function of the tip position ($j$) at fixed voltage $eV=0.7\Delta_{\rm eff}$ 
		for increasing tunneling strengths ($\Gamma/\Delta_{\rm eff}=0.1,0.3$) for different zero-energy states:  MBS 1 (a), YSR~(b), MBS 2~(c), ABS~(d), MBS 3~(e), and QMS~(f) (see Table I). In the vicinity of MBS, cases (a),(c) and (e), a flat plateau $F(j)=1$ is observed  in sharp contrast with trivial cases (b) and (d) where strong oscillations of $F$ well above $1$ are obtained. In vicinity of a QMS, case (f), $F(j)=1$ in the whole metallic region, where the Majorana wavefunctions do not overlap, however as they start overlapping in the SC region, $20<j<40$, (see \cite{SuppMat})  spatial oscillations of $F$ above 1 are found, allowing for a distinction with the isolated MBS of case (e). Dashed line: Analytical approximation obtained from the low-energy models Eq.~(\ref{eq:Fano1}) and Eq.~(\ref{eq:Fano2}).  The low-energy approximation is in excellent agreement with numerical points. The green line denotes the NS interface when it exists.
	The three insets show the DOS for the six cases considered here. 
	}
    \label{Results}
\end{figure*}

In all  cases summarized in Table I, zero energy bound states do appear well separated from other states by a gap $\Delta_{\rm eff}$. 
 This can be visualized in the single-particle density of states plotted in the insets of Fig. \ref{Results}.

In the customary weak tunneling regime, STM allows to detect the local density of states (LDoS) $\rho(j,\omega)$. However, the LDoS does not allow for a clear distinction between trivial states and MBS, because of interference or other non-universal effects 
 \cite{SuppMat}. These complications could be overcome in the strong tunneling regime, which is dominated by Andreev reflections as recently shown experimentally \cite{Ruby2015}.
In that case, it is expected that the differential conductance in the vicinity of MBS is quantized, displaying a flat spatial profile \cite{ZeroBiasPeak}. However, the oscillations of the MBS wave function ultimately spoil the conductance plateau and do not allow once again to discriminate MBS from trivial bound states \cite{Chevallier}.
In what follows, we  show that, within a rather strong tunneling regime, shot-noise STM is the right observable that allows such discrimination.

 To model the STM experiment we  define the complete Hamiltonian 
$\mathcal{H}_{tot}=\mathcal{H}_{T}+\mathcal{H}_{S}+\mathcal{H}_{tunnel},$
which includes the coupling between the tip and the sample:
\begin{align}
    \label{Htot}
    \mathcal{H}_{tunnel}&=\frac{t}{2}[\psi^\dagger_{T}\tau_z\psi_{j,S}]+\text{H.c.}.
\end{align}
Here we introduce the Nambu-spinor \hbox{$\psi_{T}=(d_{\uparrow},d_{\downarrow},d^\dagger_{\downarrow,},-d^\dagger_{\uparrow})^T$},
where the operator $d_\sigma$ annihilates an electron of spin $\sigma$ at the apex of the tip. $\mathcal{H}_T$ is the
Hamiltonian of the isolated metallic tip and $\mathcal{H}_{tunnel}$ describes the sample-tip tunneling, which is assumed purely local.
We fix $\hbar=1$ throughout.
Tunneling events are thus characterized by the energy width $\Gamma=2\pi\nu_T t^2$,
{where $\nu_T$ is} the density of states in the tip.
 The bias voltage $V$ between the tip and the
sample is taken into account in the chemical potential of the tip as $\mu_T=\mu+eV$.

%%%%%%%%%%%
{\em Current shot-noise.}
The charge current flowing from the tip is $I(j,t_1)=-e\frac{d \hat{N}_T(t_1)}{dt_1}$, where  $\hat{N}_T(t_1)$ is the number operator counting the electrons in the tip at time $t_1$ in the Heisenberg picture. In the dc regime, $I(j,t_1)=I(j)$ and 
 \begin{align}
\label{formalCurrent}
\langle I(j) \rangle=\frac{et}{2}\int \frac{d\omega}{2\pi} \Tr{G^<_{jS,T}(\omega)-G^<_{T,Sj}(\omega)},
 \end{align} where $G^<_{jS,T}(\omega)$ and $G^<_{T,Sj}(\omega)$ are local Keldysh Green's matrices at site $j$,
describing the electron hopping between the sample $S$ and the tip $T$.
 These and all the Green's functions of the tip $G^{R/A}_{T,T}$, $G^{</>}_{T,T}$ and of the substrate $G^{R/A}_{jS,jS}$, $G^{</>}_{jS,jS}$ are
 related by equations of motion with the isolated tip $g^{R/A}_{T}$, $g^{</>}_{T}$ and isolated substrate $g^{R/A}_{S}$, $g^{</>}_{S}$ bare Green's functions respectively \cite{SuppMat}. 
The shot-noise is defined as the zero-frequency limit
of the time-symmetrized current-current correlator, $S=\int d(t_1-t_2)S(t_1,t_2)$, where 
$S(t_1,t_2)=\langle\delta I(t_1)\delta I(t_2)\rangle+\langle\delta I(t_2)\delta I(t_1)\rangle$ and 
$\delta I(t_1)= I(t_1)-\langle I(t_1) \rangle$.
Using the Wick theorem \cite{SuppMat}, we find 
\begin{align}
\label{FormalSN}
S(j,eV)=&{e^2t^2}\int\frac{d\omega}{2\pi}\Tr{G^<_{T,T}(\omega)G^>_{jS,jS}(\omega)}\nonumber\\-&\Tr{G^<_{jS,T}(\omega)G^>_{jS,T}(\omega)}
+(jS\leftrightarrow T).
\end{align}  
Finally the Fano factor is given by
\begin{align}
F(j,eV)=S(j,eV)/(2e\vert I(j,eV) \vert).
\end{align}
Consequently, with the retarded uncoupled Green's functions $g^{R}_{T}$ and $g^{R}_{S}$ in hands, the charge current $I(j)$, the shot noise $S(j,eV)$ and the Fano factor $F(j,eV)$  are readily computed \cite{SuppMat}.

{\em  Fano factor tomography (FFT).}
We numerically calculated the FFT of the wire for the different configurations displayed in Table I and many values of the tunneling amplitude.
We shall present results for $\Gamma/\Delta_{\rm eff}=0.1, 0.3$, which correspond to a strong tunneling strength.
Typically, the shot-noise signal vanishes at low voltage (following the current) and rapidly saturates to finite
value for higher voltage. We thus set $eV=0.7\Delta_{\rm eff}$, which assures that the signal has reached saturation.
Since we are interested in the low-temperature quantum regime, we set
$k_BT=\frac{\Delta_{\rm eff}}{200}$.  We investigated temperature effects and our results showed that the FFT remains robust  as long as $k_B T$ is much smaller than $\Delta_{\rm eff}$  \cite{SuppMat}.
Figures. \ref{Results}()a, \ref{Results}()c and \ref{Results}(e) (which correspond to case {\it a},  {\it c}, and {\it e}, respectively, in Table I) show that the Fano factor in the vicinity of MBS states  does not significantly deviate from the Poissonian limit, $F=1$ and does not depend strongly on the tip position ($j$). This behavior is observed for any tunneling regime. Notice that for the first sites of the wire ($j<5$) $F(j)$ weakly decreases to less than one when tunneling strength increases.
 In contrast, Fig. \ref{Results}(b) (case {\it b} in Table I) shows that in vicinity of YSR bound-states, the Fano factor can reach values significantly greater than 1, oscillating strongly as a function of the tip position ($j$). This behavior is again observed for any tunneling regime and does not significantly depend on $\Gamma$. Those qualitative results are also found in the vicinity of an ABS, as plotted in Fig. \ref{Results}d 
(case {\it d} in Table I). \ps{Our results for a QMS (case {\it f)} in Table I) are shown in Fig. \ref{Results}f. This case is less discriminating  as the Fano factor oscillates above 1 only in the superconducting part of the wire where the two Majorana wave functions start overlapping.}
In the following, we provide analytical insight into these results showing that they are universally rooted to the breaking of the particle-hole symmetry of unpaired MBS.\\ 
{\em  Insights from low-energy models.}
Within our BDG formalism applied to a one-band Hamiltonian, a general zero-energy state (topological or trivial) is associated to the 4 particle-hole component wave-function $\phi_+(j)=[u_\uparrow(j),u_\downarrow(j),v_\uparrow(j),-v_\downarrow(j)]^T$ and its particle-hole partner $\phi_-(j)=\sigma_y\tau_y\mathcal{K}\phi_+(j)$, where $\mathcal{K}$ denotes the complex conjugation. 
Using the Lehmann's representation, and restricting to low-energy bound states, the sample retarded bare Green's function reads
$g^R_{S,j,j}(\omega)\approx \frac{\phi_+(j)\phi_+^\dagger(j)+\phi_-(j)\phi_-^\dagger(j)}{\omega+i0^+}.$~Under the general assumption that the sample Hamiltonian is real \footnote{We numerically check that our results remain valid for complex hamiltonians, \cite{SuppMat}.},
$u_\sigma(j)$ and $v_\sigma(j)$ can be safely chosen as real numbers. 
By using the equations of motion  \cite{SuppMat}, we are able to obtain  exact expressions for both the current and the shot-noise.
In the general case of finite temperature and finite voltage, these expressions turn out to be lengthy and not insightful \cite{SuppMat}.
However, in the zero-temperature limit and saturated voltage regime, $eV\gg\Gamma_j$, with $\Gamma_j=\Gamma\sum_\sigma( \vert u_\sigma(j)\vert^2+\vert v_\sigma(j)\vert^2)$, we obtain  
\begin{align}
    S(j)&\simeq{8e^2\Gamma} \frac{(\sum_\sigma u_\sigma^2)(\sum_\sigma v_\sigma^2)[(\sum_\sigma u_\sigma^2)^2+(\sum_\sigma v_\sigma^2)^2]}{(\sum_\sigma u_\sigma^2+\sum_\sigma v_\sigma^2)^3}, \label{eq:S}  \\
    I(j)&\simeq {2e} \Gamma\frac{(\sum_\sigma u_\sigma^2)(\sum_\sigma v_\sigma^2)}{(\sum_\sigma u_\sigma^2+\sum_\sigma v_\sigma^2)}, \label{eq:I} \\
    F(j)&\simeq 1+ \left(\frac{\sum_\sigma (u_\sigma^2-v_\sigma^2)}{\sum_\sigma (u_\sigma^2+v_\sigma^2)}\right)^2= 1+\delta_{\rm ph}^2(j),\label{eq:Fano1}
\end{align}
where $\delta_{\rm ph}(j)$ denotes the local particle-hole asymmetry.
Consequently $F(j)$ does not depend on the tunneling strength and $1\leq F(j)$. 
Furthermore, and most importantly, we prove that $F(j)-1$ is in direct correspondence with $\delta_{\rm ph}(j)$ and can thus display strong spatial variations. Nonzero $\delta_{\rm ph}(j)$ obviously requires finite overlap of Majorana components, we thus expect $F$ to be sensitive to MBS overlap~\cite{SuppMat}.
Indeed, we show in Figs. \ref{Results}(b), \ref{Results}(d) and \ref{Results}(f) that the low-energy approximation of Eq. \eqref{eq:Fano1} is in excellent quantitative agreement with the numerical results.

In sharp contrast, a perfectly isolated MBS has a {local} particle-hole symmetry for any position,
imposing strong constraints on the spatial dependence of $F(j)$. 
An isolated MBS is described by the retarded Green's function,
$g^R_{S,j,j}(\omega)=\frac{\phi_M(j)\phi_M^\dagger(j)}{\omega+i0^+}$
with $\phi_M(j)=(u_\uparrow,u_\downarrow,u_\downarrow^*,-u_\uparrow^*)^T$. 
Setting $T=0$, the  shot-noise and current can be obtained exactly and have a compact form even for finite voltage.
The resulting Fano factor reads
\footnote{This result only relies on the property of the Majorana wave function.}
\begin{align}
\label{eq:Fano2}
F(j,eV)&= 1-\frac{\Gamma_j}{\arctan(\frac{eV}{\Gamma_j})}\frac{eV}{(eV)^2+\Gamma_j^2}.
\end{align}
Therefore, in the saturated regime, $eV\gg\Gamma_j $, $F(j)\approx 1$ and the Fano factor does not depend on $j$. Moreover, the leading order corrections in $\frac{\Gamma_j}{eV}$ reads $  F(j)\simeq 1-\frac{2\Gamma_j}{\pi eV}$, showing that the  Fano factor in the vicinity of a MBS weakly decreases with increasing tunneling strength $\Gamma$ and
takes values lower than 1, in agreement with the numerical curves displayed in Figs. \ref{Results}(a), \ref{Results}(c) and \ref{Results}(e).

As any zero-energy Dirac fermions can be written as a pair of Majorana fermions \cite{Kitaev2000}, the
  breaking of the particle-hole symmetry, producing the large overshooting of the Fano factor above one, can be interpreted in terms of
 a spatial overlap of the two Majorana wavefunctions. This must be contrasted with
  the particle-hole symmetry preserved {in a spatially} well isolated MBS.
  We can directly prove this by analyzing how the Fano factor departs from $F\sim 1$ by increasing the overlap of the Majorana wavefunctions in the nanowire model (see \cite{SuppMat}).
These arguments are general and rely only on the existence of a zero-energy bound state in an effective single-band Hamiltonian. 
 Although the FFT  does not provide any direct signature of the topology  of the bulk substrate, it is of great practical interest to select those  MBS,  which do not overlap and are therefore resilient to local noise which makes them good candidates for quantum computation \cite{Prada2020}.

{\em Conclusions and perspectives.}
 We show that the recently developed
 STM shot-noise techniques can provide a key discerning tool MBS from other  zero-energy fermionic states.
  In particular,  we evidence that the Fano factor strongly oscillates spatially for  ABS and YSR bound states
	around the impurity location, with amplitudes greatly exceeding one, i.e the Poissonian limit. This must be sharply contrasted with the behavior of the MBS Fano factor, which barely
deviates from  one. These sharp differences have a universal character which is rooted in the intrinsic
    particle-hole symmetry of the MBS wavefunction.
		Such signature in the FFT in vicinity of a zero-bias conductance peak thus constitutes  an additional necessary condition to  identify non-overlapped MBS although it does not directly access the topology of the bulk.	
  Our results should foster further experimental developments of STM shot-noise experiments in the field of Majorana fermions and, more generally, in topological matter.

{\em Acknowledgments.} We acknowledge {fruitful} discussions and exchange with M. Aprili, D. Chevallier, F. Massee, A. Mesaros, and A. Palacios-Morales.

\bibliography{Biblio_Noise2}

\end{document}

\title{{Supplementary Material: Discriminating Majorana by tunneling shot-noise tomography}}% Force line breaks with \\

\author{Vivien Perrin }
\affiliation{%
Universit\'e Paris-Saclay, CNRS, Laboratoire de Physique des Solides, 91405, Orsay, France
}%

\author{Marcello Civelli }
\affiliation{%
Universit\'e Paris-Saclay, CNRS, Laboratoire de Physique des Solides, 91405, Orsay, France
}%
\author{Pascal Simon }
\affiliation{%
Universit\'e Paris-Saclay, CNRS, Laboratoire de Physique des Solides, 91405, Orsay, France
}%

\date{\today}% It is always \today, today,
             %  but any date may be explicitly specified

\begin{abstract}
  In this supplementary material, we give further technical details about the Keldysh Green's functions, the general derivation of the current and noise, its application to the calculation of noise for trivial and MBS bound states, and results about the LDoS and differential conductance tomography.
\end{abstract}

%\keywords{Suggested keywords}%Use showkeys class option if keyword
                              %display desired
\maketitle

%\begin{widetext}
\newcommand{\beginsupplement}{%
        \setcounter{section}{0}
        \renewcommand{\thesection}{S~\Roman{section}}
        \setcounter{equation}{0}
        \renewcommand{\theequation}{S\arabic{equation}}
        \setcounter{figure}{0}
        \renewcommand{\thefigure}{S\arabic{figure}}
        \setcounter{table}{0}
        \renewcommand{\thetable}{S\arabic{table}}%%}
    }
\beginsupplement 
\section{Keldysh Green's functions in Nambu space}
In the main text (MT) we defined four-component Nambu-spinors of the STM tip $\psi_T$ and substrate $\psi_{l,S}$. Namely,
\begin{align}
    \psi_{l,S}&=(\psi_{e,\uparrow,l,S},\psi_{e,\downarrow,l,S},\psi_{h,\uparrow,l,S},\psi_{h,\downarrow,l,S})^T=(c_{\uparrow,l},c_{\downarrow,l},c^\dagger_{\downarrow,l},-c^\dagger_{\uparrow,l})^T,\\
    \psi_{T}&=(\psi_{e,\uparrow,T},\psi_{e,\downarrow,T},\psi_{h,\uparrow,T},\psi_{h,\downarrow,T})^T=(d_{\uparrow},d_{\downarrow},d^\dagger_{\downarrow},-d^\dagger_{\uparrow})^T.
\end{align}
Nambu-spinors are made of electron and holes operators, thus $\psi_{l,S}^*= (c^\dagger_{\uparrow,l},c^\dagger_{\downarrow,l},c_{\downarrow,l},-c_{\uparrow,l})^T$ and $\psi_{l,S}$ are not independent. Namely Nambu-spinors are invariant under particle-hole transformation ,
\begin{align}
    \psi_{l,S}=\sigma_y.\tau_y.\mathcal{K}\psi_{l,S},
\end{align}
where $\mathcal{K}$ is the complex conjugation operator. The same relation holds for $\psi_T$.
It is possible to use Keldysh Green's functions for the Nambu-spinors. They are thus 4x4 in matrices Nambu-space whose components are defined by:
\begin{align}
    G^R_{\alpha,\tau,\sigma;\beta,\tau',\sigma'}(t,t')&=-i\theta(t-t')\langle\lbrace\psi_{\alpha,\tau',\sigma'}(t),\psi^\dagger_{\beta,\tau',\sigma'}(t')\rbrace\rangle,\\
    G^A_{\alpha,\tau,\sigma;\beta,\tau',\sigma'}(t,t')&=-i\theta(t'-t)\lbrace\psi_{\alpha,\tau',\sigma'}(t),\psi^\dagger_{\beta,\tau',\sigma'}(t')\rbrace,\\
    G^<_{\alpha,\tau,\sigma;\beta,\tau',\sigma'}(t,t')&=i\langle\psi^\dagger_{\beta,\tau',\sigma'}(t')\psi_{\alpha,\tau',\sigma'}(t)\rangle,\\
    G^>_{\alpha,\tau,\sigma;\beta,\tau',\sigma'}(t,t')&=-i\langle\psi_{\alpha,\tau',\sigma'}(t)\psi^\dagger{\beta,\tau',\sigma'}(t')\rangle.
\end{align}
Where $\tau,\tau'$ are electron-hole indices, $\sigma,\sigma'$ spin ones, $\alpha,\beta$ denotes other indices (spatial index $j$ and lead index $S,T$) and $\lbrace A,B \rbrace$ denotes the anti-commutator of operators $A$ and $B$. In DC regime, the previous Green's functions depends only on the time difference $(t-t')$ , it is then possible to decompose Green's functions in terms of  their Fourier components,
\begin{align}
    G^\gamma_{\alpha,\beta}(t,t')=\int\frac{d\omega}{2\pi}e^{-i\omega(t-t')}G^\gamma_{\alpha,\beta}(\omega),\\
    G^\gamma_{\alpha,\beta}(\omega)=\int d(t-t') G^\gamma_{\alpha,\beta}(t,t')e^{i\omega(t-t')}.
\end{align}
\section{Current and shot-noise}
First let us derive Eq. 3. The current operator $I(j,t)$ , defined in the MT reads,
\begin{align}
    I(j,t_1)&=-\frac{e}{\hbar}\partial_{t_1}\lbrace\sum_\sigma d^\dagger_\sigma(t_1)d_\sigma(t_1)\rbrace=\frac{iet}{\hbar}\sum_\sigma\lbrace d^\dagger_\sigma(t_1)c_{\sigma,j}(t_1)-c^\dagger_{\sigma,j}(t_1)d_{\sigma}(t_1)\rbrace.
\end{align}
Assuming DC regime, a straightforward substitution of Eq. 1,Eq. 2 in Eq. 10  leads to,
\begin{align}
    \langle I(j,t_1)\rangle&=\frac{iet}{2\hbar}\sum_{\tau,\sigma}\lbrace\psi_{T,\tau,\sigma}^\dagger(t_1)\psi_{j,S,\tau,\sigma}(t_1)-\psi_{j,S,\tau,\sigma}^\dagger(t_1)\psi_{T,\tau,\sigma}(t_1)\rbrace\nonumber\\&=\frac{et}{2\hbar}\int \frac{d\omega}{2\pi} \Tr{G^<_{jS,T}(\omega)-G^<_{T,jS}(\omega)}.
\end{align}
Now we turn to the derivation of Eq. 4.
Since the Hamiltonian is quadratic, Wick's theorem can be used and one obtains,
\begin{align}
    S(t_1,t_2)=&\langle\delta I(t_1)\delta I(t_2)\rangle+(t_1\leftrightarrow t_2)\nonumber\\
    =&-\frac{e^2t^2}{4\hbar^2}\lbrace\sum_{\tau_1,\sigma_1,\tau_2,\sigma_2}\langle\psi^\dagger_{T,\sigma_1,\tau_1}(t_1)\psi_{S,\sigma_1,\tau_1}(t_1)\psi^\dagger_{T,\sigma_2,\tau_2}(t_2)\psi_{S,\sigma_2,\tau_2}(t_2)\rangle\nonumber\\
    &-\sum_{\tau_1,\sigma_1,\tau_2,\sigma_2}\langle\psi^\dagger_{T,\sigma_1,\tau_1}(t_1)\psi_{S,\sigma_1,\tau_1}(t_1)\psi^\dagger_{S,\sigma_2,\tau_2}(t_2)\psi_{T,\sigma_2,\tau_2}(t_2)+(S\leftrightarrow T)\rangle\rbrace+(t_1\leftrightarrow t_2)\nonumber\\
    &-2\langle I(t_1)\rangle\langle I(t_2)\rangle\nonumber\\
    =&\frac{e^2t^2}{2\hbar^2}\sum_{\sigma_1,\tau_1,\sigma_2,\tau_2}\lbrace\langle\psi^\dagger_{T,\sigma_1,\tau_1}(t_1)\psi_{\sigma_2,\tau_2}(t_2)\rangle\langle\psi_{S,\sigma_1,\tau_1}(t_1)\psi^\dagger_{\sigma_2,\tau_2}(t_2)\rangle\nonumber\\
    &-\langle\psi^\dagger_{T,\sigma_1,\tau_1}(t_1)\psi_{\sigma_2,\tau_2}(t_2)\rangle\langle\psi_{S,\sigma_1,\tau_1}(t_1)\psi^\dagger_{\sigma_2,\tau_2}(t_2)\rangle+(S\leftrightarrow T)\rbrace+(t_1\leftrightarrow t_2)\nonumber\\
    =&\frac{e^2t^2}{2\hbar^2}\Tr{G^<_{T,T}(t_2,t_1)G^>_{S,S}(t_1,t_2)-G^<_{S,T}(t_2,t_1)G^>_{S,T}(t_1,t_2)+(S\leftrightarrow T)}+(t_1\leftrightarrow t_2).
\end{align}
Where we omit the site index $j$ for brevity and used Eq.~S3. Finally, after a straightforward Fourier transformation, the  current shot-noise reads,
\begin{align}
    S(j)=\frac{e^2t^2}{\hbar^2}\int\frac{d\omega}{2\pi}&\Tr{G^<_{T,T}(\omega)G^>_{jS,jS}(\omega)}\nonumber-\Tr{G^<_{jS,T}(\omega)G^>_{jS,T}(\omega)}\nonumber+(jS\leftrightarrow T).
\end{align}
We have then recovered Eq. 4.
\section{Equations of motion}
\label{EOM}
As mentioned in the MT equations of motion allows us to compute all Green's functions from isolated ones. Indeed, using the closed set of equation of motions (EOM) :
\begin{align}
    G^{R/A}_{jS,jS}(\omega)=&g^{R/A}_{S,j,j}(\omega)+t^2g^{R/A}_{S,j,j}(\omega)\tau_z g^{R/A}_{T}(\omega)\tau_zG^{R/A}_{jS,jS}(\omega),\\
    G^{R/A}_{T,T}(\omega)=&g^{R/A}_{T}(\omega)+t^2g^{R/A}_{T}(\omega)\tau_z g^{R/A}_{S,j,j}(\omega)\tau_zG^{R/A}_{T,T}(\omega),\\
    G^{</>}_{T,T}(\omega)=&g^{</>}_{T}(\omega)+t^2g^{</>}_{T}(\omega)\tau_z g^{A}_{S,j,j}(\omega)\tau_zG^{A}_{T,T}(\omega)\\&+t^2g^{R}_{T}(\omega)\tau_z g^{</>}_{S,j,j}(\omega)\tau_zG^{A}_{T,T}(\omega)+t^2g^{R}_{T}(\omega)\tau_z g^{R}_{S,j,j}(\omega)\tau_zG^{</>}_{T,T}(\omega)\nonumber,\\
    G^{</>}_{jS,jS}(\omega)=&g^{</>}_{jS}(\omega)+t^2g^{</>}_{S,j,j}(\omega)\tau_z g^{A}_{T}(\omega)\tau_zG^{A}_{jS,jS}(\omega)\\&+t^2g^{R}_{S,j,j}(\omega)\tau_z g^{</>}_{T}(\omega)\tau_zG^{A}_{jS,jS}(\omega)+t^2g^{R}_{S,j,j}(\omega)\tau_z g^{R}_{T}(\omega)\tau_zG^{</>}_{jS,jS}(\omega)\nonumber,\\
    G^{</>}_{jS,T}(\omega)=&tG^{</>}_{jS,jS}(\omega)\tau_z g^A_{T}(\omega)+tG^{R}_{jS,jS}(\omega)\tau_z g^{</>}_{T}(\omega)\nonumber,\\
    G^{</>}_{T,jS}(\omega)=&tg^{</>}_{T}(\omega)\tau_z G^A_{jS,jS}(\omega)+tg^{</>}_{T}(\omega)\tau_z G^A_{jS,jS}(\omega)\nonumber.
\end{align}
All Green's functions can be obtained from the knowledge of isolated Green's functions. The isolated metallic Tip is described by the following Green's functions,
\begin{align}
    g_T^{R/A}(\omega)=&\mp i\pi\nu_T{\sigma}_0\otimes{\tau}_0,\\
 g_T^<(\omega)=2i\pi\nu_T &[n_F(\omega^-){\sigma}_0\otimes\frac{{\tau}_0+{\tau}_z}{2}\nonumber+n_F(\omega^+){\sigma}_0\otimes\frac{{\tau}_0-{\tau}_z}{2}],\\
    g_T^>(\omega)=-2i\pi\nu_T &[\bar{n}_F(\omega^-){\sigma}_0\otimes\frac{{\tau}_0+{\tau}_z}{2}\nonumber+\bar{n}_F(\omega^+){\sigma}_0\otimes\frac{{\tau}_0-{\tau}_z}{2}].
\end{align}
Where $n_F(\omega)$ is the Fermi-Dirac distribution function, $\bar{n}_F(\omega)=1-n_F(\omega)$, $\omega^\pm=\omega\pm eV$ and $\nu_T$ is the density of states in the Tip at the Fermi level.
In consequence once $g^{R}_{S,j,j}(\omega)$ in hand, we are able to solve this set of EOM and compute current and current shot-noise.
\section{Densities of state and differential conductances}
\begin{figure}[htb!!]
    \centering
    \includegraphics[width=15cm]{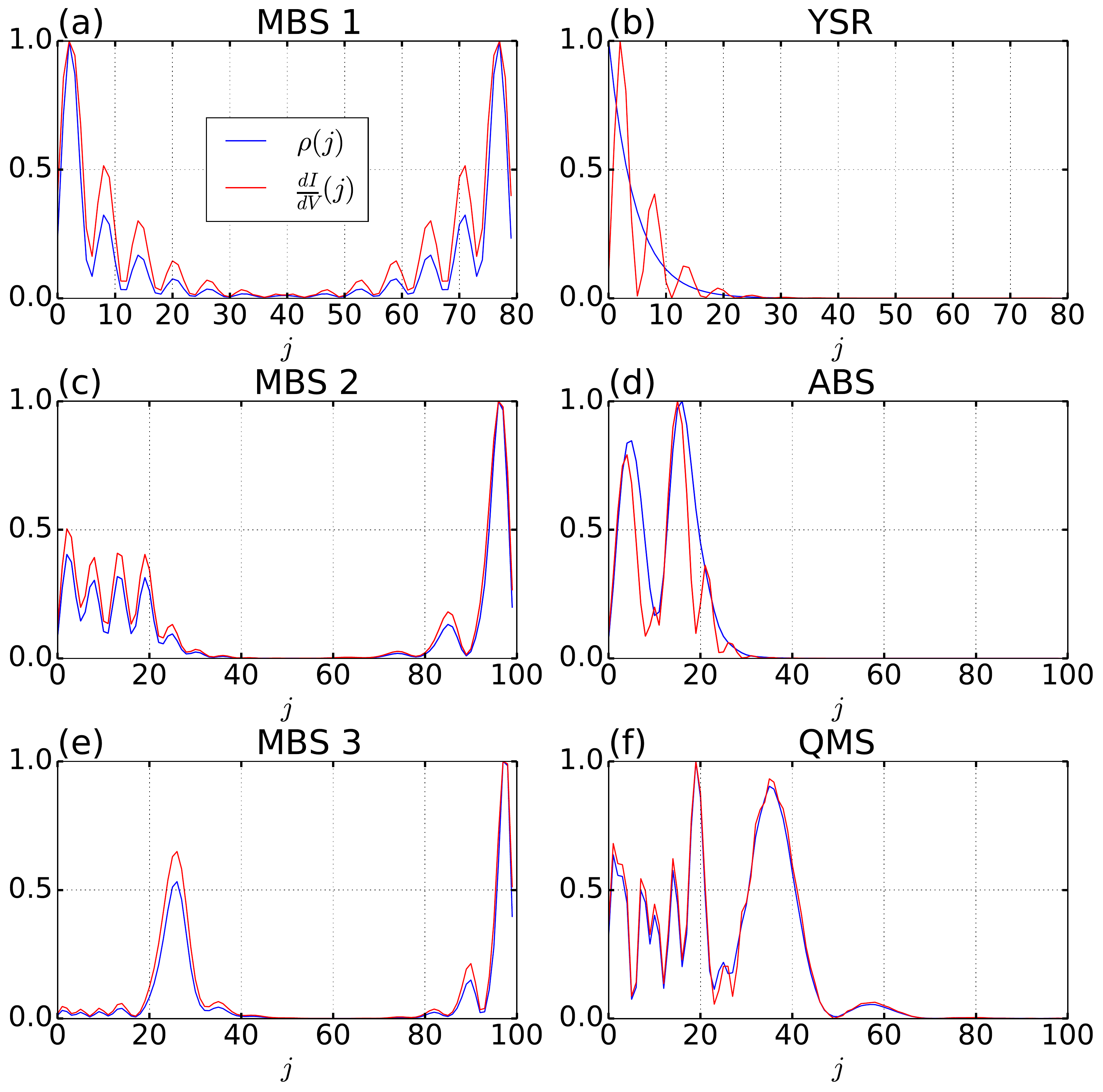}
    \caption{(Solid blue line) Density of states at zero-energy $\rho(\omega=0,j)=\frac{-1}{2\pi}{\rm Im}{\rm Tr}\lbrace g^R_{S,j,j}(0)\rbrace$  normalized to its maximum for the 6 parameters set of Table I. (Solid red line)Zero-bias differential conductance at zero-energy $\frac{dI}{dV}(V=0)$  normalized to its maximum for the 4 parameters set of Table I. The tunneling strength is $\Gamma/\Delta_{eff}=0.2$ and $k_BT=\Delta_{eff}/200$.}
    \label{DoS}
\end{figure}
We mentioned in the MT that the local density of states $\rho(j,\omega)$ does not allow for a clear distinction between MBS and other trivial zero-energy states, since there is no qualitative difference between the LDoS of MBS and other trivial states. On Fig.~\ref{DoS}, we plotted the LDoS obtained for the 6 different parameters set of Table I. First we remark that contrary to the MBS cases, the YSR LDoS shown on Fig.~\ref{DoS}(b) is smooth and does not oscillates. The absence of spatial oscillations in the LDoS of zero-energy YSR is not due to fine tuning of parameters but a general feature of these states\footnote{Indeed, relying on a continuum approximation, we can show that the electron and hole components of the zero-energy YSR wave function $u_\uparrow(j)$ and $v_\downarrow(j)$ possess the same exponential decay and spatially oscillate in perfect phase opposition \cite{Rusinov1969,Menard2015,Vardan2016} and therefore the LDoS proportional to $u_\uparrow^2(j)+v_\downarrow^2(j)$ does not oscillate.}. It may thus tempting to use the LDoS to discriminate MBS from trivial YSR  zero-energy. However the MBS LDoS can be made smoother by changing the parameter of the wire, and the ABS LDoS also shows spatial oscillations, forbidding the use of the oscillating LDoS as a criterion to distinguish trivial states from MBS. Moreover, we also mentioned that we assumed a strong tunneling regime where Andreev reflections dominates and discard single-particle processes. As a consequence, the differential conductance at zero-bias, shown on Fig.~\ref{DoS}, is not proportional to the LDoS. In all considered cases, the differential conductance at zero-bias strongly oscillates as a function of space and we do not observe any saturation plateau in vicinity of MBS. The differential conductance is thus not able to distinguish MBS from trivial states here.
\section{Details for the low-energy approximation }
\label{Analytics}
\subsection{Zero-energy trivial bound-state}
As mentioned in the MT, a single zero-energy bound-state is described by the green's function:
\begin{align}
\label{ABSGF}
g^R_{S,j,j}(\omega)=&\frac{\phi_+(j)\phi^\dagger_+(j)+\phi_-(j)\phi^\dagger_-(j)}{\omega+i0^+},
\end{align} 
where $\phi_+(j)=(u_\uparrow(j),u_\downarrow(j),v_\downarrow(j),-v_\uparrow(j))^T$ is the wave-function associated to the zero-energy bound-state and $\ket{\phi_-(j)}=\sigma_y\tau_y\mathcal{K}\phi_+(j)$ its particle-hole symmetric counter-part. To simplify further the problem, we assume that the substrate Hamiltonian can be made real, hence $u_\sigma,v_\sigma$ are real functions of space. Inserting this expression in the previous set of EOM, and assuming $T=0$, $eV\gg\Gamma_j$ we obtain 
\begin{align}
    I(j,eV)&\simeq e\int\frac{d\omega}{2\pi} \frac{8 \Gamma ^2 \left(4 \omega^2 \left(u_\downarrow^2 \left(2 v_\downarrow^2+v_\uparrow^2\right)+2 u_\downarrow u_\uparrow v_\downarrow v_\uparrow+u_\uparrow^2 \left(v_\downarrow^2+2 v_\uparrow^2\right)\right)+\Gamma ^2  (u_\uparrow v_\downarrow-u_\downarrow v_\uparrow)^2 \alpha_j\right)}{8 \Gamma ^2 \omega^2 4\Gamma_j^2(u_\downarrow v_\downarrow + u_\uparrow v_\uparrow)^2+\Gamma ^4 \alpha(j)^2+16 \omega^4}, \\
    \alpha_j&=\left((u_\downarrow-v_\downarrow)^2+(u_\uparrow-v_\uparrow)^2\right)\left((u_\downarrow+v_\downarrow)^2+(u_\uparrow+v_\uparrow)^2\right),\\
    S(j,eV)&\simeq e^2\int\frac{d\omega}{2\pi} 24\Gamma^2\frac{ A\omega^6+B\omega^4\Gamma^2 + C\omega^2\Gamma^4+D\Gamma^6}{[8 \Gamma ^2 \omega^2 4\Gamma_j^2(u_\downarrow v_\downarrow + u_\uparrow v_\uparrow)^2+\Gamma ^4 \alpha_j^2+16 \omega^4]^2},\\
    A&=64 \lbrace2 u_\downarrow^2 v_\downarrow^2+u_\downarrow^2 v_\uparrow^2+2 u_\downarrow u_\uparrow v_\downarrow v_\uparrow+u_\uparrow^2 v_\downarrow^2+2 u_\uparrow^2 v_\uparrow^2\rbrace,\\
    B&=16\lbrace4 u_\downarrow^6 v_\downarrow^2+3 u_\downarrow^6 v_\uparrow^2+2 u_\downarrow^5 u_\uparrow v_\downarrow v_\uparrow+11 u_\downarrow^4 u_\uparrow^2 v_\downarrow^2+10 u_\downarrow^4 u_\uparrow^2 v_\uparrow^2-8 u_\downarrow^4 v_\downarrow^4-14 u_\downarrow^4 v_\downarrow^2 v_\uparrow^2+2 u_\downarrow^4 v_\uparrow^4+4 u_\downarrow^3 u_\uparrow^3 v_\downarrow v_\uparrow\\&-4 u_\downarrow^3 u_\uparrow v_\downarrow^3 v_\uparrow-36 u_\downarrow^3 u_\uparrow v_\downarrow v_\uparrow^3+10 u_\downarrow^2 u_\uparrow^4 v_\downarrow^2+11 u_\downarrow^2 u_\uparrow^4 v_\uparrow^2-14 u_\downarrow^2 u_\uparrow^2 v_\downarrow^4+20 u_\downarrow^2 u_\uparrow^2 v_\downarrow^2 v_\uparrow^2-14 u_\downarrow^2 u_\uparrow^2 v_\uparrow^4+4 u_\downarrow^2 v_\downarrow^6\nonumber\\&+11 u_\downarrow^2 v_\downarrow^4 v_\uparrow^2+10 u_\downarrow^2 v_\downarrow^2 v_\uparrow^4+3 u_\downarrow^2 v_\uparrow^6+2 u_\downarrow u_\uparrow^5 v_\downarrow v_\uparrow-36 u_\downarrow u_\uparrow^3 v_\downarrow^3 v_\uparrow-4 u_\downarrow u_\uparrow^3 v_\downarrow v_\uparrow^3\nonumber\\&+2 u_\downarrow u_\uparrow v_\downarrow^5 v_\uparrow+4 u_\downarrow u_\uparrow v_\downarrow^3 v_\uparrow^3+2 u_\downarrow u_\uparrow v_\downarrow v_\uparrow^5+3 u_\uparrow^6 v_\downarrow^2+4 u_\uparrow^6 v_\uparrow^2+2 u_\uparrow^4 v_\downarrow^4-14 u_\uparrow^4 v_\downarrow^2 v_\uparrow^2-8 u_\uparrow^4 v_\uparrow^4+3 u_\uparrow^2 v_\downarrow^6\\&+10 u_\uparrow^2 v_\downarrow^4 v_\uparrow^2\nonumber+11 u_\uparrow^2 v_\downarrow^2 v_\uparrow^4+4 u_\uparrow^2 v_\uparrow^6\rbrace\nonumber,\\
    C&=4\lbrace 2 v_\downarrow^2 u_\downarrow^{10}+3 v_\uparrow^2 u_\downarrow^{10}-2 u_\uparrow v_\downarrow v_\uparrow u_\downarrow^9-8 v_\downarrow^4 u_\downarrow^8+4 v_\uparrow^4 u_\downarrow^8+11 u_\uparrow^2 v_\downarrow^2 u_\downarrow^8+14 u_\uparrow^2 v_\uparrow^2 u_\downarrow^8-20 v_\downarrow^2 v_\uparrow^2 u_\downarrow^8-56 u_\uparrow v_\downarrow v_\uparrow^3 u_\downarrow^7\\&\nonumber+8 u_\uparrow v_\downarrow^3 v_\uparrow u_\downarrow^7-8 u_\uparrow^3 v_\downarrow v_\uparrow u_\downarrow^7+12 v_\downarrow^6 u_\downarrow^6+2 v_\uparrow^6 u_\downarrow^6-36 u_\uparrow^2 v_\downarrow^4 u_\downarrow^6-12 u_\uparrow^2 v_\uparrow^4 u_\downarrow^6+48 v_\downarrow^2 v_\uparrow^4 u_\downarrow^6+24 u_\uparrow^4 v_\downarrow^2 u_\downarrow^6\\&\nonumber+26 u_\uparrow^4 v_\uparrow^2 u_\downarrow^6\nonumber+42 v_\downarrow^4 v_\uparrow^2 u_\downarrow^6+16 u_\uparrow^2 v_\downarrow^2 v_\uparrow^2 u_\downarrow^6+84 u_\uparrow v_\downarrow v_\uparrow^5 u_\downarrow^5-24 u_\uparrow v_\downarrow^3 v_\uparrow^3 u_\downarrow^5-104 u_\uparrow^3 v_\downarrow v_\uparrow^3 u_\downarrow^5-12 u_\uparrow v_\downarrow^5 v_\uparrow u_\downarrow^5\\&-40 u_\uparrow^3 v_\downarrow^3 v_\uparrow u_\downarrow^5\nonumber-12 u_\uparrow^5 v_\downarrow v_\uparrow u_\downarrow^5-8 v_\downarrow^8 u_\downarrow^4+4 v_\uparrow^8 u_\downarrow^4+42 u_\uparrow^2 v_\downarrow^6 u_\downarrow^4+48 u_\uparrow^2 v_\uparrow^6 u_\downarrow^4-12 v_\downarrow^2 v_\uparrow^6 u_\downarrow^4-44 u_\uparrow^4 v_\downarrow^4 u_\downarrow^4\\&-44 u_\uparrow^4 v_\uparrow^4 u_\downarrow^4-44 v_\downarrow^4 v_\uparrow^4 u_\downarrow^4\nonumber-102 u_\uparrow^2 v_\downarrow^2 v_\uparrow^4 u_\downarrow^4+26 u_\uparrow^6 v_\downarrow^2 u_\downarrow^4+24 u_\uparrow^6 v_\uparrow^2 u_\downarrow^4-36 v_\downarrow^6 v_\uparrow^2 u_\downarrow^4+132 u_\uparrow^2 v_\downarrow^4 v_\uparrow^2 u_\downarrow^4\\&+72 u_\uparrow^4 v_\downarrow^2 v_\uparrow^2 u_\downarrow^4-56 u_\uparrow v_\downarrow v_\uparrow^7 u_\downarrow^3\nonumber-104 u_\uparrow v_\downarrow^3 v_\uparrow^5 u_\downarrow^3-24 u_\uparrow^3 v_\downarrow v_\uparrow^5 u_\downarrow^3-40 u_\uparrow v_\downarrow^5 v_\uparrow^3 u_\downarrow^3+272 u_\uparrow^3 v_\downarrow^3 v_\uparrow^3 u_\downarrow^3\\&-40 u_\uparrow^5 v_\downarrow v_\uparrow^3 u_\downarrow^3+8 u_\uparrow v_\downarrow^7 v_\uparrow u_\downarrow^3-24 u_\uparrow^3 v_\downarrow^5 v_\uparrow u_\downarrow^3\nonumber-104 u_\uparrow^5 v_\downarrow^3 v_\uparrow u_\downarrow^3-8 u_\uparrow^7 v_\downarrow v_\uparrow u_\downarrow^3+2 v_\downarrow^{10} u_\downarrow^2+3 v_\uparrow^{10} u_\downarrow^2\\&-20 u_\uparrow^2 v_\downarrow^8 u_\downarrow^2-20 u_\uparrow^2 v_\uparrow^8 u_\downarrow^2+14 v_\downarrow^2 v_\uparrow^8 u_\downarrow^2+48 u_\uparrow^4 v_\downarrow^6 u_\downarrow^2\nonumber+42 u_\uparrow^4 v_\uparrow^6 u_\downarrow^2+26 v_\downarrow^4 v_\uparrow^6 u_\downarrow^2+16 u_\uparrow^2 v_\downarrow^2 v_\uparrow^6 u_\downarrow^2-12 u_\uparrow^6 v_\downarrow^4 u_\downarrow^2\\&-36 u_\uparrow^6 v_\uparrow^4 u_\downarrow^2+24 v_\downarrow^6 v_\uparrow^4 u_\downarrow^2+72 u_\uparrow^2 v_\downarrow^4 v_\uparrow^4 u_\downarrow^2+132 u_\uparrow^4 v_\downarrow^2 v_\uparrow^4 u_\downarrow^2\nonumber+14 u_\uparrow^8 v_\downarrow^2 u_\downarrow^2+11 u_\uparrow^8 v_\uparrow^2 u_\downarrow^2+11 v_\downarrow^8 v_\uparrow^2 u_\downarrow^2\\&+16 u_\uparrow^2 v_\downarrow^6 v_\uparrow^2 u_\downarrow^2-102 u_\uparrow^4 v_\downarrow^4 v_\uparrow^2 u_\downarrow^2+16 u_\uparrow^6 v_\downarrow^2 v_\uparrow^2 u_\downarrow^2-2 u_\uparrow v_\downarrow v_\uparrow^9 u_\downarrow-8 u_\uparrow v_\downarrow^3 v_\uparrow^7 u_\downarrow\nonumber+8 u_\uparrow^3 v_\downarrow v_\uparrow^7 u_\downarrow-12 u_\uparrow v_\downarrow^5 v_\uparrow^5 u_\downarrow\\&-40 u_\uparrow^3 v_\downarrow^3 v_\uparrow^5 u_\downarrow-12 u_\uparrow^5 v_\downarrow v_\uparrow^5 u_\downarrow-8 u_\uparrow v_\downarrow^7 v_\uparrow^3 u_\downarrow-104 u_\uparrow^3 v_\downarrow^5 v_\uparrow^3 u_\downarrow-24 u_\uparrow^5 v_\downarrow^3 v_\uparrow^3 u_\downarrow+8 u_\uparrow^7 v_\downarrow v_\uparrow^3 u_\downarrow\nonumber-2 u_\uparrow v_\downarrow^9 v_\uparrow u_\downarrow\\&-56 u_\uparrow^3 v_\downarrow^7 v_\uparrow u_\downarrow\nonumber+84 u_\uparrow^5 v_\downarrow^5 v_\uparrow u_\downarrow-56 u_\uparrow^7 v_\downarrow^3 v_\uparrow u_\downarrow-2 u_\uparrow^9 v_\downarrow v_\uparrow u_\downarrow+3 u_\uparrow^2 v_\downarrow^{10}+2 u_\uparrow^2 v_\uparrow^{10}+4 u_\uparrow^4 v_\downarrow^8\\&-8 u_\uparrow^4 v_\uparrow^8\nonumber+11 u_\uparrow^2 v_\downarrow^2 v_\uparrow^8+2 u_\uparrow^6 v_\downarrow^6+12 u_\uparrow^6 v_\uparrow^6+24 u_\uparrow^2 v_\downarrow^4 v_\uparrow^6-36 u_\uparrow^4 v_\downarrow^2 v_\uparrow^6+4 u_\uparrow^8 v_\downarrow^4-8 u_\uparrow^8 v_\uparrow^4+26 u_\uparrow^2 v_\downarrow^6 v_\uparrow^4\\&-44 u_\uparrow^4 v_\downarrow^4 v_\uparrow^4+42 u_\uparrow^6 v_\downarrow^2 v_\uparrow^4\nonumber+3 u_\uparrow^{10} v_\downarrow^2+2 u_\uparrow^{10} v_\uparrow^2+14 u_\uparrow^2 v_\downarrow^8 v_\uparrow^2-12 u_\uparrow^4 v_\downarrow^6 v_\uparrow^2+48 u_\uparrow^6 v_\downarrow^4 v_\uparrow^2-20 u_\uparrow^8 v_\downarrow^2 v_\uparrow^2\rbrace\nonumber,\\
    D&=\left((u_\downarrow-v_\downarrow)^2+(u_\uparrow-v_\uparrow)^2\right)^2 (u_\uparrow v_\downarrow-u_\downarrow v_\uparrow)^2 \left(u_\downarrow^2+u_\uparrow^2-v_\downarrow^2-v_\uparrow^2\right)^2 \left((u_\downarrow+v_\downarrow)^2+(u_\uparrow+v_\uparrow)^2\right)^2.
\end{align}
Where we omitted spatial index of $u_\sigma(j), v_\sigma(j)$ for brevity. Performing the integrals lead to,
\begin{align}
    S(j)&\simeq{4e^2\Gamma} \frac{(\sum_\sigma u_\sigma^2)(\sum_\sigma v_\sigma^2)((\sum_\sigma u_\sigma^2)^2+(\sum_\sigma v_\sigma^2)^2)}{(\sum_\sigma u_\sigma^2+\sum_\sigma v_\sigma^2)^3},\\
    I(j)&\simeq {2e} \Gamma\frac{(\sum_\sigma u_\sigma^2)(\sum_\sigma v_\sigma^2)}{(\sum_\sigma u_\sigma^2+\sum_\sigma v_\sigma^2)},\\
    F(j)&\simeq 1+ \lbrace\frac{\sum_\sigma |u|_\sigma^2-|v|_\sigma^2}{\sum_\sigma |u|_\sigma^2+|v|_\sigma^2}\rbrace^2= 1+\delta_{\rm ph}^2(j).
\end{align}.
which are Eq. 6, 7 and 8.
\subsection{Isolated MBS}
\label{Shot-MBS}
The case of a perfect isolated MBS is simpler. Indeed, due to the intrinsic particle-hole symmetry shot-noise and current an be obtained analytically without any additional reality assumption. The local retarded green's function describing an isolated MBS reads\cite{ZeroBiasPeak}, 
\begin{align}
    g^R_{S,j,j}(\omega)=\frac{\phi_M(j)\phi^\dagger_M(j)}{\omega},
\end{align}
where $\phi_M(j)=(u_\uparrow(j),u_\downarrow(j),u_\downarrow(j)^*,-u_\uparrow(j)^*)^T$ is the wave-function of the MBS which satisfy the pseudo-reality condition, $\phi_M(j)=\sigma_y\tau_y\mathcal{K}\phi_M(j)$. Injecting this expression in our set of EOM, we obtain the results ,
\begin{align}
    I(j,eV)=&\frac{e}{2\pi}\int \frac{\Gamma_j^2}{\omega^2+\Gamma_j^2}[n_F(\omega^-)-n_F(\omega^+)],\\
    S(j,eV)=&\frac{2e^2\Gamma_j}{\pi}\int \frac{n_F(\omega^-)}{(\omega^2+\Gamma_j^2)^2}[\Gamma_j^2\bar{n}_F(\omega^-)+\omega^2\bar{n}_F(\omega^+)]+(eV\leftrightarrow-eV).
\end{align}
In agreement with previous works \cite{KTLaw,T.Martin}.
In the $T=0$ limit the integrals can be performed analytically and we obtain ,
\begin{align}
    I(j,eV)=&\frac{e\Gamma_j}{\pi}\arctan(\frac{eV}{\Gamma_j}),\\
    S(j,eV)=&\frac{2e^2\Gamma_j}{\pi}[\arctan(\frac{eV}{\Gamma_j})-\frac{eV\Gamma_j}{(eV)^2+\Gamma_j^2}],\\
    F(j,eV)=&1-\frac{eV\Gamma_j}{[(eV)^2+\Gamma_j^2]\arctan(\frac{eV}{\Gamma_j})]}.
\end{align}
Consequently, taking the saturated limit, $eV\gg\Gamma_j$ we obtained,
\begin{align}
    F(j)\approx 1-\frac{2\Gamma_\xi}{\pi eV}+\mathcal{O}(\frac{1}{(eV)^2}).
\end{align}
We have then recovered Eq. 10.
\newpage
\section{Zero-energy Dirac fermions as a pair of Majorana fermions}

Obviously any zero-energy Dirac fermion $c$, can be interpreted as a pair of zero-energy Majorana fermions $\gamma_A$,$\gamma_B$, such that,
    $c=e^{i\varphi}\frac{\gamma_A+i\gamma_B}{\sqrt{2}}$,
with, $\gamma^\dagger_A=\gamma_A$, 
    $\gamma^\dagger_B=\gamma_B$,
    $\lbrace\gamma_B,\gamma_A\rbrace=0$
and $\varphi$ an arbitrary phase.
\subsection{Majorana wavefunctions in the BdG formalism}
This result is easily seen in our BdG formalism. Indeed, a zero-energy Dirac fermions is associated to a pair $\phi_+$ and $\phi_-$ of zero-energy degenerate  eigenvectors of the first quantized BdG Hamiltonian ($H_{BdG}$). Consequently, any linear combination of those eigenvectors, is still an eigenvector of $H_{BdG}$. We can then define $\phi_A(j)=\frac{e^{i\varphi}\phi_+(j)+e^{-i\varphi}\phi_-(j)}{\sqrt{2}}$ and $\phi_B(j)=-i \frac{e^{i\varphi}\phi_+(j)-e^{-i\varphi}\phi_-(j)}{\sqrt{2}}$ which are zero-energy eignvectors of $H_{BdG}$ and fulfills the pseudo-reality condition, $\phi_{A/B}(j)=\tau_y\sigma_y\mathcal{K}\phi_{A/B}(j)$. Thus $\phi_{A/B}=(u_{\uparrow,A/B},u_{\downarrow,A/B},u^*_{\downarrow,A/B},-u^*_{\uparrow,A/B})^T$ are Majorana wavefunctions. For a Rashba nano-wire in the topological phase, the Majorana wavefunctions associated to the zero-energy state have separated spatial supports and then the non-local zero-energy fermion can be interpreted as two unpaired MBS each one localized at one edge of the wire.
\subsection{Spatial variations of $F>1$ as a consequence of Majorana overlap} 
In general, there is no reason for the Majorana wave-functions associated with a zero-energy femion to be spatially separated, and $\phi_A(j)$,$\phi_B(j)$ can be simultaneously non-zero in vicinity of the bound-states. Hence  Eq.~S29 reads \begin{align}F\simeq 1-\frac{\left(u_{\downarrow,B} u_{\downarrow,A}^*-u_{\downarrow,A} u_{\downarrow,B}^*+u_{\uparrow,B} u_{\uparrow,A}^*-u_{\uparrow,A} u_{\uparrow,B}^*\right)^2}{\left(\left| u_{\downarrow,A}\right| ^2+\left| u_{\downarrow,B}\right| ^2+\left| u_{\uparrow,A}\right| ^2+\left| u_{\uparrow,B}\right| ^2\right)^2}.\end{align}.
It is then obvious that in presence of a finite overlap of the MBS wavefunctions, $F(j)$ is no longer restricted to 1. However in the case of isolated MBS, since $\phi_A(j)$ is non-zero only where $\phi_B(j)$ vanishes, $F(j)=1$. This establish a direct link between the absence of Majoranas overlap and spatial plateau in the Fano factor.

\subsection{Majorana wave-functions of the MT}

\begin{figure}[H]
    \centering
    \includegraphics[height=13cm]{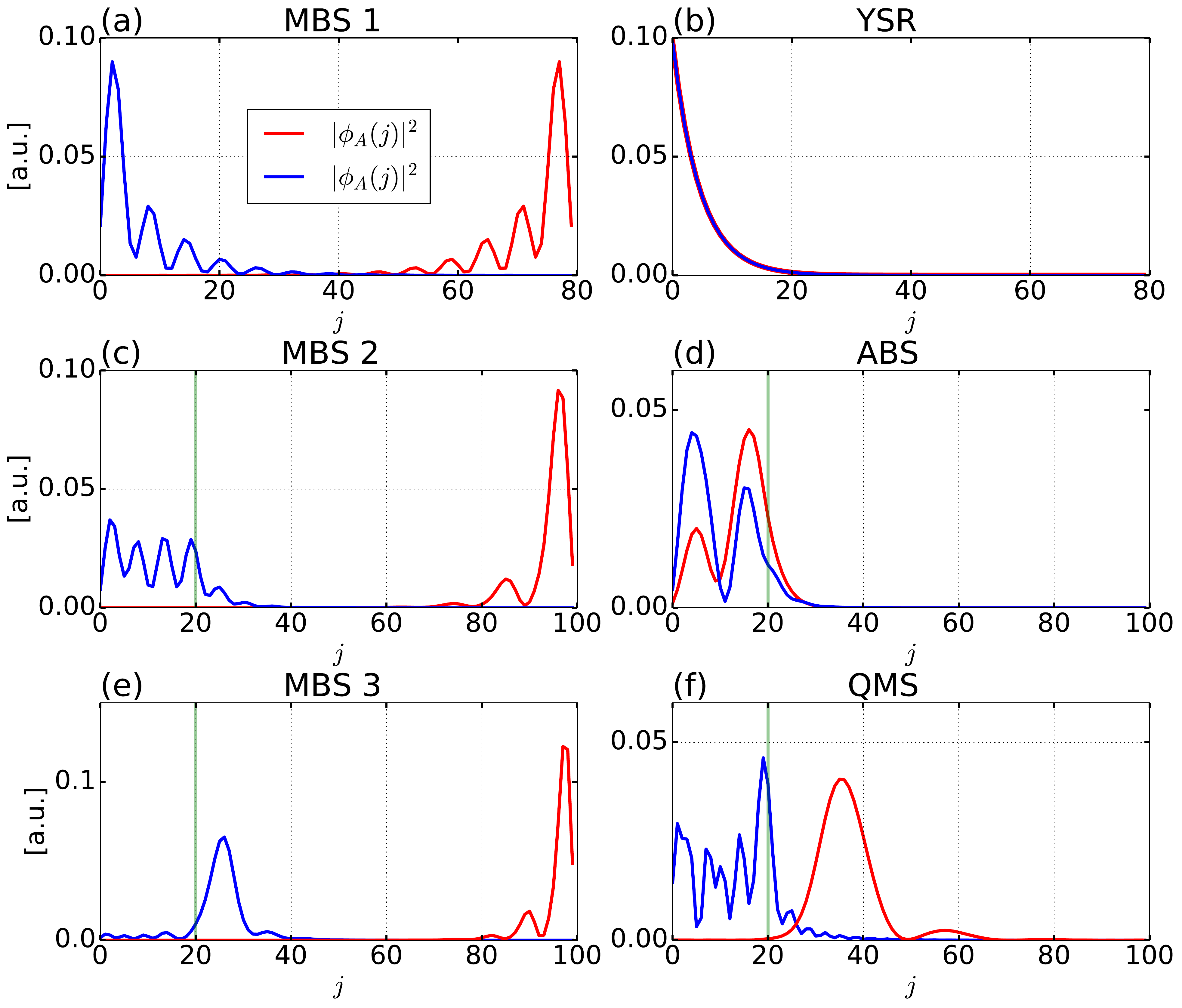}
    \caption{Local weight of the 2  Majorana wavefunctions $|\phi_A(j)|^2=\sum_\sigma|u_{A,\sigma}|^2$ (red) and $|\phi_B(j)|^2=\sum_\sigma|u_{B,\sigma}|^2$ (blue) associated with the zero-energy Dirac fermions in the 6 different cases of Table~1. The green line denote the NS interface when there exists one. In all MBS cases (a),(c)and (e) the wavefunctions are well separated in space and do not overlap each other. In the trivial YSR case, (b), the wavefunctions are perfectly overlapped. In the ABS case , (d), the wavefunctions  overlap on the whole left edge. To finish the case of a QMS is of particular interest since the Majorana wavefunctions only weakly ovelaps but are still located at the same edge of the wire. All those results, are in agreement with the loaction of spatial variations of $F$ above $1$ on Fig~2. }
    \label{MajoranaWFS}
\end{figure}

On figure Fig. \ref{MajoranaWFS} we show the spatial profile of the Majorana wavefunctions associate to the 6 sets of parameters in the MT. From this figure it is clear that regions where $F(j)$ deviates from 1 correspond to regions where Majorana wavefunctions significantly overlap. Indeed, for all MBS cases (a),(c)and (e) the wavefunctions are well separated in space and do not overlap each other, in agreement with the results of Fig~2 where the $F$ is pinned to one i vicinity of the bound-states. In the  YSR case, (b), the wavefunctions are perfectly overlapped, in agreement with the observation of spatial oscillations of $F$ extending over the whole left edge of the wire (Fig~2). In the ABS case , (d), the wavefunctions fully overlap , again in agreement with the observation of $F\neq1$ both in the whole normal region of the wire. To finish the case of a QMS is of particular interest since the Majorana wavefunctions only weakly ovelaps but are still located at the same ege of the wire. The origin of the partial separation of the wavefunctions is rooted in the smoothness of the potential barrier $V(l)$\cite{ABSZoologyReview,Vuik2019}. Thus in the whole normal region the wavefunction are not overlapped,in agreement with the plateau $F=1$ observed n the whole normal region on Fig~2. However in the SC region, the wavefunctions are weakly overlapped, for sites 20<j<40, which correspond to the regions where we observed spatial variations of $F$ above 1 on Fig~2. This suggests that Fano factor tomography can be useful to etect weak overlap of Majorana wavfunctions and distinguish unpaired MBS from QMBS with reminiscent overlap. 
\newpage
\begin{figure}[H]
    \centering
    \includegraphics[height=15cm]{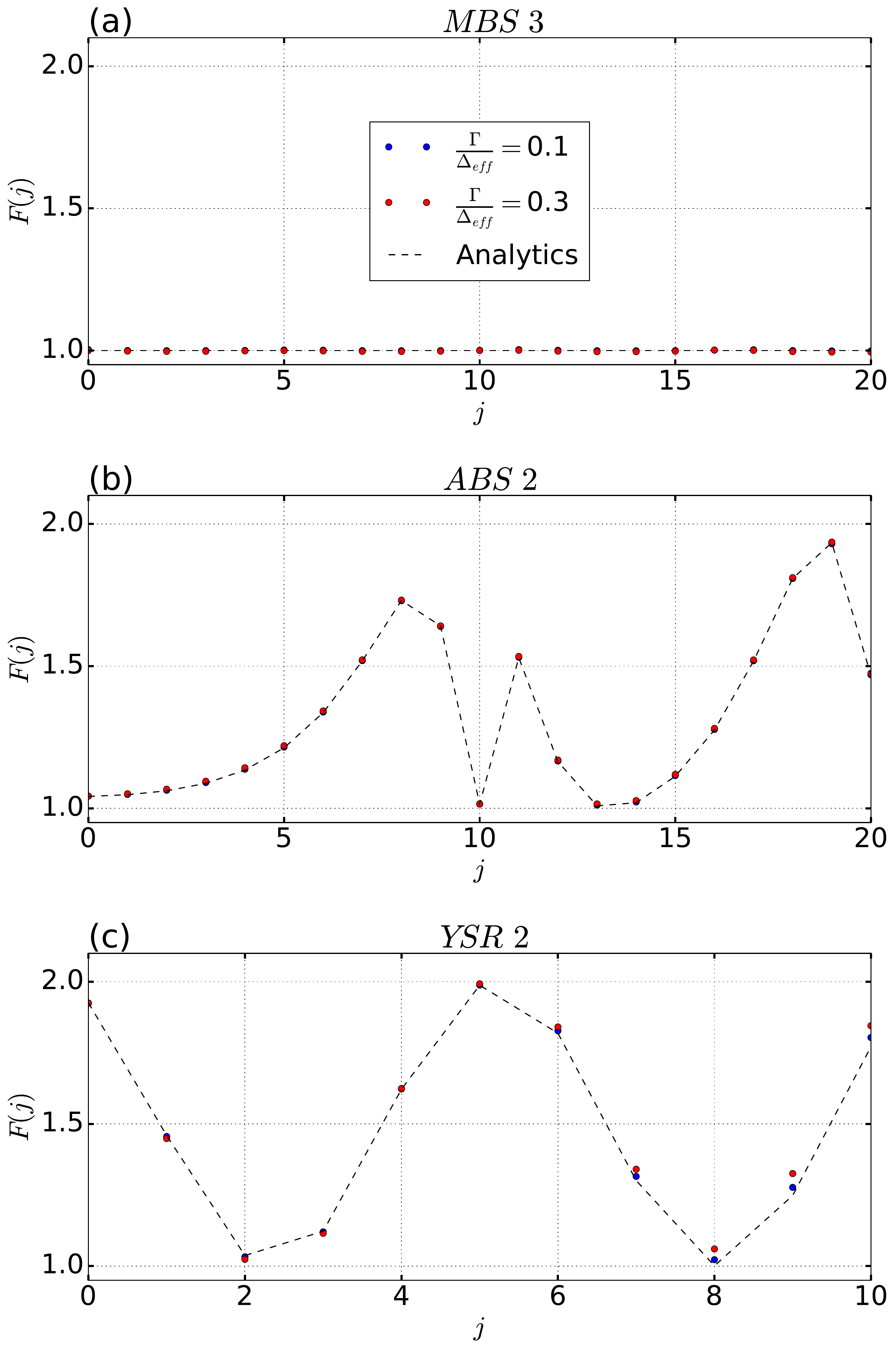}
    \caption{(a,b,c) Square amplitude ( in arbitrary units) of the Majorana wave-functions associated to the zero-energy fermionic states for $V_Z=1.5,3.05,3.65$ respectively. In case (a), the MBS overlap at the edges of the wire ($j\leq10$) is zero. In cases (b) and (c) the Majorana wave-functions are not confined to one edge of the wire and there exits a finite overlap at each ends of the wire. In all cases, the energy of the state is zero and the overlap does not lift the MBS degeneracy.\\(d) Spectrum of the wire as a function of the Zeeman coupling $V_Z$. In all figures the parameters are ($\mu=0.5,t_w=10,\Delta=1,\alpha=1.2,N=60$)}
    \label{SpectumOverlap}
\end{figure}
\subsection{Tomography of overlapping topological MBS}
Due to the finite length of the wire, the MBS states localized at the edges of Rashba nanowire in the topological phase can also present a finite overlap at the edges.

\begin{figure}[H]
    \centering
    \includegraphics[width=15cm]{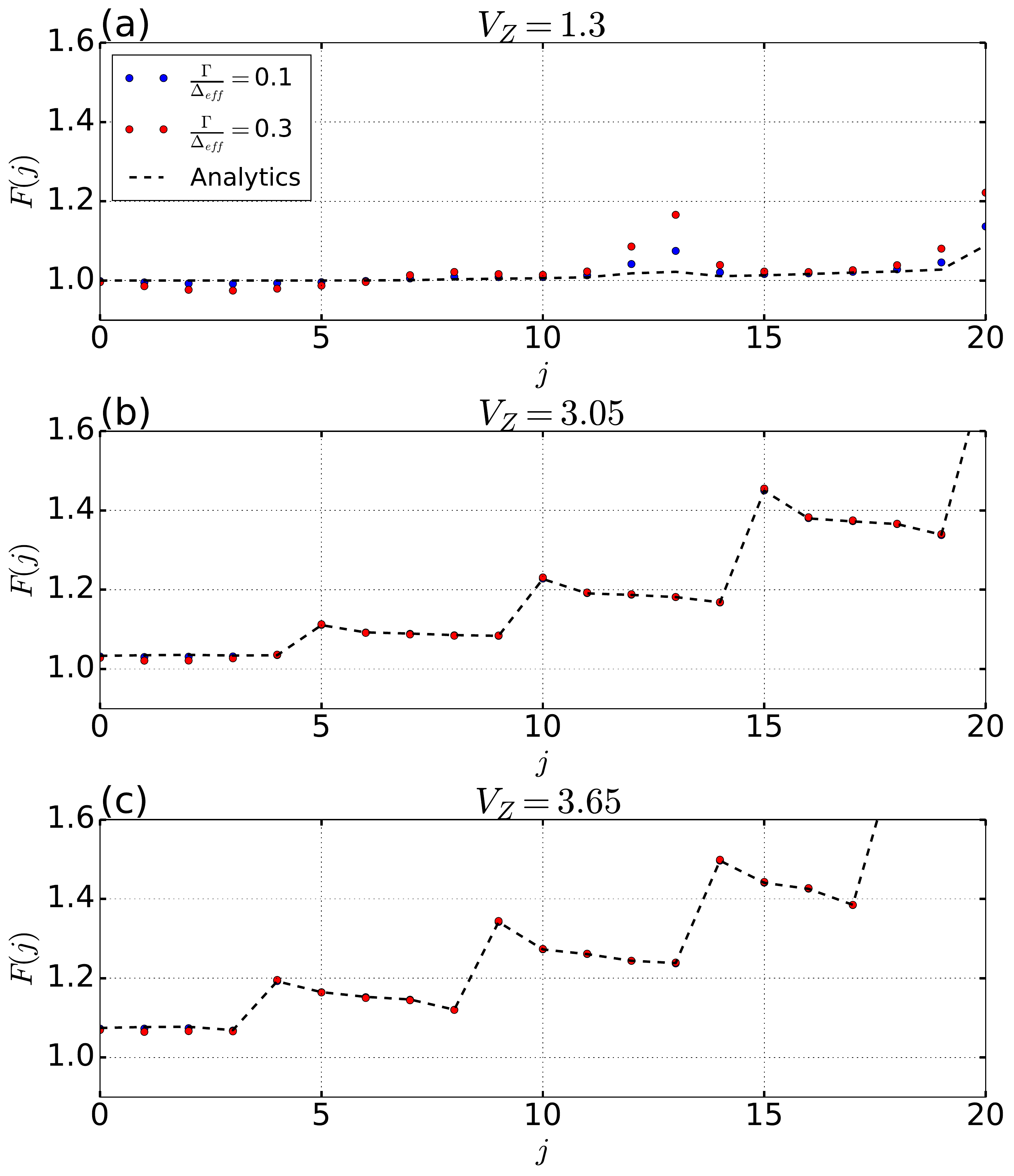}
    \caption{ Fano factor $F$ as a function of the tip position, $j$ for $V_Z=1.5$(a), $V_Z=3.05$(b) and $V_Z=3.5$(c). Other parameters are set to ($\mu=0.5,t_w=10,\Delta=1,\alpha=1.2,N=6$). In case (a) the absence of Majorana wave-functions overlap leads to the observation of a plateau $F=1$ for the first sites of the wire where the MBS is localized ($j\leq 10$).  In case (b) and (c) the presence of a finite overlap of the MBS at the edges washes out the Fano plateau $F=1$ and MBS overlap is responsible for finite particle-hole assymetry of the zero-energy states (as it can observed from the dashed dark line), increasing $F(j)$ above 1 even at the edges ($j\leq10$). In all cases, the analytical approximation of Eq.~S29/S36 (black dashed line) is in quantitative agreement with the numerical results. For all cases, the voltage and temperature are $eV=0.7\Delta_{\rm eff}$ and $k_BT=\Delta_{\rm eff}/200$. }
    \label{OverlapResults}
\end{figure}

 Generally, finite overlap of Majorana wavefunctions leads to a finite energy of the MBS, but there exists fine-tuning point for the Zeeman coupling where zero-energy MBS exists with a strong overlap( see Fig.~\ref{SpectrumOverlap}). In such a case, the finite overlap of the MBS wave-functions should be detected by Fano tomography according to Eq. S29. Fig.~\ref{OverlapResults} shows that the plateau $F=1$ observed in absence (Panel (a)) of overlap disappear in the presence of finite overlap (panels (b) and (c)) which is responsible for an increase in $F$ above 1. \\Indeed, case (a) the absence of Majorana wave-functions overlap leads to the observation of a plateau $F=1$ for the first sites of the wire where the MBS is localized ($j\leq 10$) For position further in the bulk , ($j>10$), the Majorana wave-function $\phi_A$ is weak (see Fig.~\ref{SpectumOverlap}(a)) and non-universal contributions are responsible for the increase in $F(j)$ above 1. Those effects are of two types. First due the weak weight of $\phi_A$ the perfect isolated MBS description of the zero-energy fermion is no longer valid. Indeed, one can see from the dashed dark line that $\delta_{ph}(j)^2$ is non-zero for such position. Such contribution are then finite-size effects. Second, due the weak weight of the MBS wave-functions $\phi_A$ and $\phi_B$ at positions $j > 10$ the non-resonant Andreev reflections on bulk states becomes dominant charge transfer process and increase $F$. This is especially true for $j=13$ where MBS wave-functions are almost exactly vanishing.\footnote{ However,the current at this position is also infinitesimally small and then the Fano-factor may not be experimentally accessible here and one should observe  a plateau extending from $j=0$ to $j=20$.}\\On the contrary, in cases (b) and (c), the presence of a finite overlap of the MBS at the edges washes out the Fano plateau $F=1$ and MBS overlap is responsible for finite particle-hole assymetry of the zero-energy states (as it can observed from the dashed dark line), increasing $F(j)$ above 1 for all positions.\\
 To finish, in presence of finite overlap of the MBS wave-functions, Eq.~S29/S36 still describes quantitatively the exact numerical results. 
\newpage
\section{Fano factor Tomography with complex Hamiltonians:}
\begin{figure}[H]
    \centering
    \includegraphics[height=10cm]{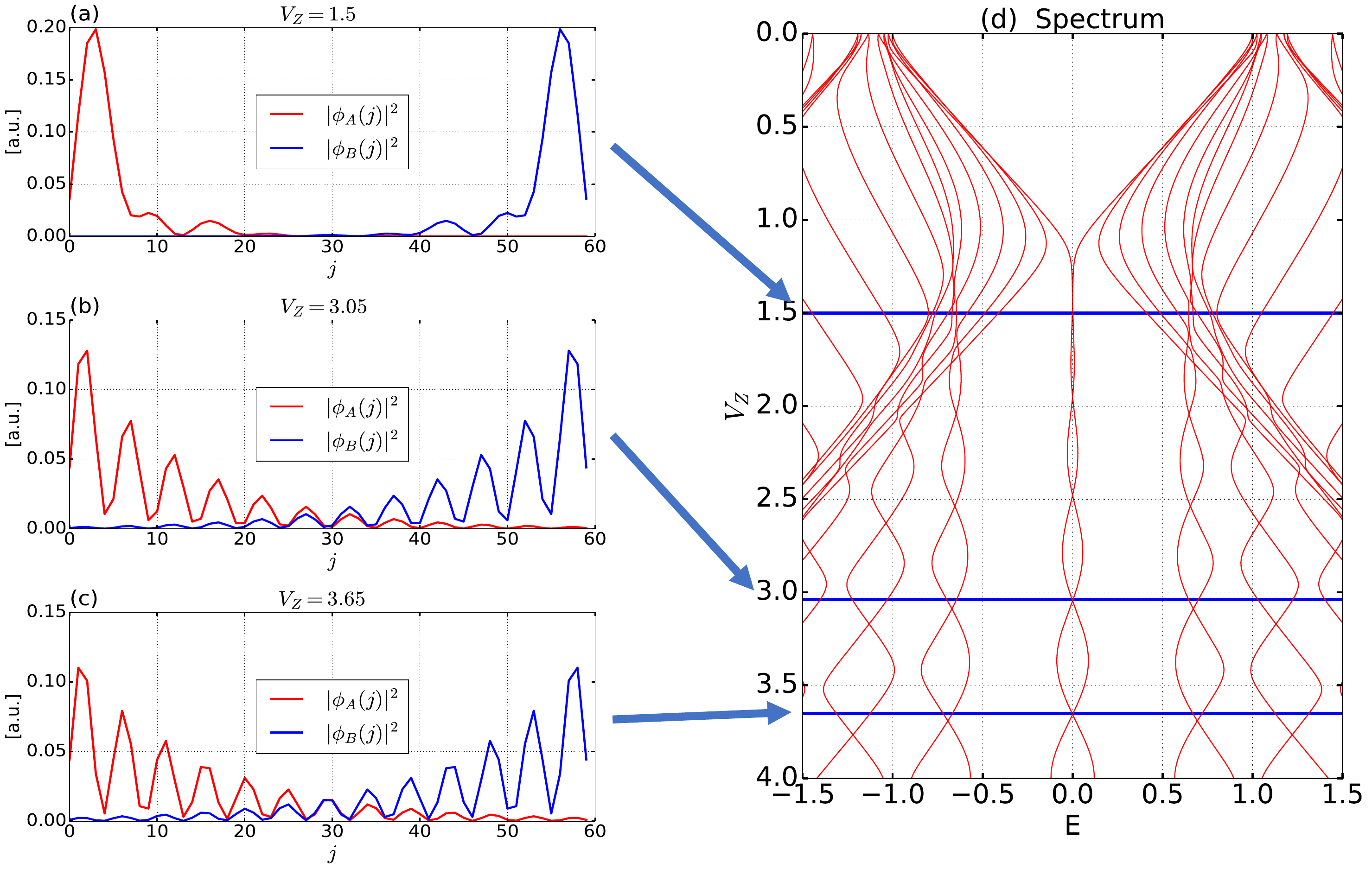}  
    \caption{(Dots) Fano factor ($F$) as a function of the tip position ($j$) at fixed voltage $eV=0.7\Delta_{\rm eff}$ and temperature $k_BT=\frac{\Delta_{\rm eff}}{200}$ for increasing tunneling strengths ($\Gamma$) and wire configurations 'MBS 4'(a), 'ABS 2'(b), 'YSR 2'(c) . In vicinity of MBS  , case (a) a flat plateau $F(j)=1$ is observed  in sharp contrast with trivial cases (b) and (c) where strong oscillations of $F$ well above $1$ are observed.(Dashed line) Analytical approximation obtained from the low-energy models Eq.~S36 and Eq.~S29.  The low-energy approximation is in excellent agreement with numerical points. The weak disagreement for $j>5$ in panel (c) is due to the contribution of bulk state which is neglected in Eq.~S29 and become important when the tip moves away from the YSR location.
	}
    \label{ComplexResults}
\end{figure}

In the previous section, we used the assumption that the Hamiltonian, giving birth to trivial bound-states can be made real to obtain Eq.~S29. Here we show that this relation is still valid for complex Hamiltonians by numerical investigations of three additional set of parameters for the wire. To that end, we investigate both the possibility for the external magnetic field giving rise to the Zeeman splitting to lie in the $(x,y)$ plane and the possibility for the magnetic impurity to be polarized along the $y$-axis. The Hamiltonian of the wire then rads,
\begin{align}
    \mathcal{H}_{S}=&\frac{1}{2}
    \sum_{l=0}^{N-1}
    \psi^\dagger_{l,S}[(2t-\mu)\tau_z+\Delta(l)\tau_x+V_Z(\cos\theta\sigma_x+\sin\theta\sigma_y)]\psi_{l,S}\nonumber\\&
		+\frac{1}{2}\sum_{l=0}^{N-2}\psi^\dagger_{l+1,S}[-{t_w}\tau_z-i\alpha\sigma_y\tau_z]\psi_{l,S} + h.c.\nonumber\\
    &-\frac{1}{2}\psi^\dagger_{0,S} J\sigma_y \psi_{0,S}~.
\end{align}
We investigate 3 specific set of parameters({\it a)}, {\it b)},and {\it c)}) detailed in Table.~S1. In all cases, the Hamiltonian contains complex terms and can not be made real. In case {\it a)} the wire supports MBS at its ends. In case {\it b)} it supports a zero-energy ABS localized at its left end. In case {\it c)} it supports a zero-energy YSR at its left end.
\begin{table}[H]
\label{TableParams}
\begin{center}
\begin{tabular}{|c|c|c|c|c|c|c|c|c|c|c|c|c|}
%%%%%%%%%%%%%%%%%%%%%%%%%%%%%%%%%%%%
   \hline
          &  Configuration & $\mu$ & $t_w$ & $\Delta$ & $V_Z$ &$\theta$& $\alpha$ & $J$ &$N$&$N_N$&$\Delta_{\rm eff}$\\\hline
 {\it a)} &    MBS 4 & 0.5 & 10 & 1  & 1.5 & $\frac{\pi}{6}$&2&0 & 100&20&0.26 \\
 {\it b)} &    ABS 2 & 0.5 & 10 & 1  & 0.26 & $\frac{\pi}{6}$&2&0 & 100&20&0.27 \\
 {\it c)} &    YSR 2  & 0.5 & 10 & 0.6  & 0 & 0&1.2&11.23 & 80&0&0.6\\
 \hline \end{tabular}
\end{center}
\label{TableComplex}
\caption{Table summarizing 3 additional parameter sets where the wire Hamiltonian can not be made real. %The parameters have been chosen such that
   $\Delta_{\rm eff}$ is the effective gap separating the zero-energy bound-state from other states.} 
\end{table} Again we numerically computed $F(j)$ as function of tip's position, results of the Fano factor tomography are pesented on Fig.~\ref{ComplexResults}. Fig.~\ref{ComplexResults}(a), shows that even in presence of complex terms in the Hamiltonian, $F(j)$ still shows a plateau $F(j)=1$ in vicinity of a MBS. This was expected since Eq.~S36 relies only on the form of MBS wave-function without any other additional assumptions about the Hamiltonian. Fig.~\ref{ComplexResults}(b) shows that even in the presence of complex terms in the Hamiltonian Eq.~S29 obtained with the assumption of a real Hamiltonian, is still valid, indeed, the black dashed line corresponding to Eq.~S29 is in perfect quantitative agreement with the numerical results ( color dots) for the first sites of the wire. Moreover, $F(j)$ still significantly exceed 1 in vicinity of trivial states.
\section{Temperature Effects}
\begin{figure}[H]
    \centering
    \includegraphics[width=0.9\linewidth]{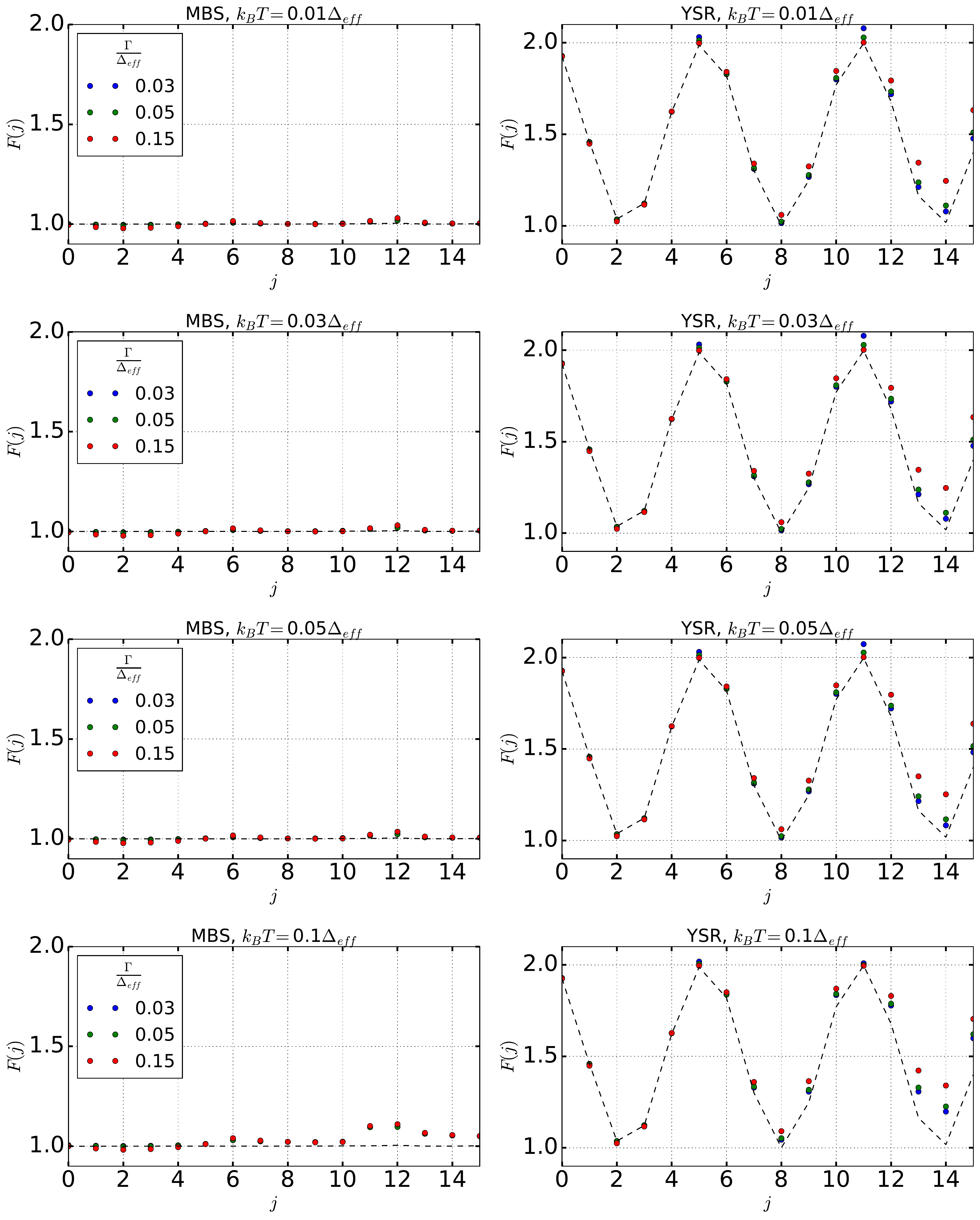}
    \caption{$F(j)$ as function of the tip position for a nanowire in configuration {\it a} (left-column) and {\it b} (right-column of Tble~1 at various temperatures. Each row corresponds to a fixed temperature.Colored dots re the numerical simulation results and the black solid-dashed line correspond to the $T=0$ and low-energy approximation of Eq.~S29. The results show that there is no significant temperature dependence for temperature up to $0.1\Delta_{\rm eff}$ where weak quantitative deviations are observed. All simulations have been performed setting $eV=0.7\Delta_{\rm eff}$.}
    \label{TempDependence}
\end{figure}

The simulations presented in the MT was performed within the extremely low temperature regime, $k_BT=\frac{\Delta_{\rm eff}}{200}$. Moreover the low-energy analytical approach  was performed in the pure $T=0$ where the current fluctuations purely come from quantum origins. It is thus legitimate to study how the low-temperature results of the MT are modified when temperature increases. To answer the question, we performed numerical simulation of the Fano tomography on  nano-wire in vicinity of a true MBS (configuration {\it a)} of Table~1) and of a YSR (configuration {\it b)} of Table~1). Our results are presented on Fig~\ref{TempDependence} shows that independently of the tunneling energy width $\Gamma$, the low-energy results are not significantly modified by thermal effects for temperature $k_BT<0.1\Delta{\rm eff}$. When temperature increases further, thermal contribution to shot-noise are responsible for an increase of the factor which weakly spoils the Poissonian plateau in vicinity of MBS. However, the qualitative behavior of $F(j)$ in vicinity of the zero-energy state ($j<12$) is still well described by our analytical approximation of Eq.~S29 (black-dashed line) for all temperatures studied here. Temperature effects are  more pronounced in the tails of the zero-energy states  where the quantum noise contribution is weak and easily dominated by temperature contributions. Our results suggests that as long as $k_B T\ll \Delta_{\rm eff}$ and $k_B T < eV-\Delta_{\rm eff}$ (such that bulk contributions do not play a key role), the low-temperature results are not modified. Consequently, the proposed Fano tomography would be relevant in current state  of the art experimental set-ups based on nanowire hybrid structure. Indeed in \cite{Mourik2012} authors reported $kT \simeq 100 mK$ and $\Delta_{\rm eff}\simeq 200 \mu eV$, leading to $\frac{k_BT}{\Delta_{\rm eff}}\simeq 0.05 $, where the low-temperature regime is still valid.
\section{Particle/hole local weights of the zero-energy states}

\begin{figure}[H]
    \centering
    \includegraphics[width=0.9\linewidth]{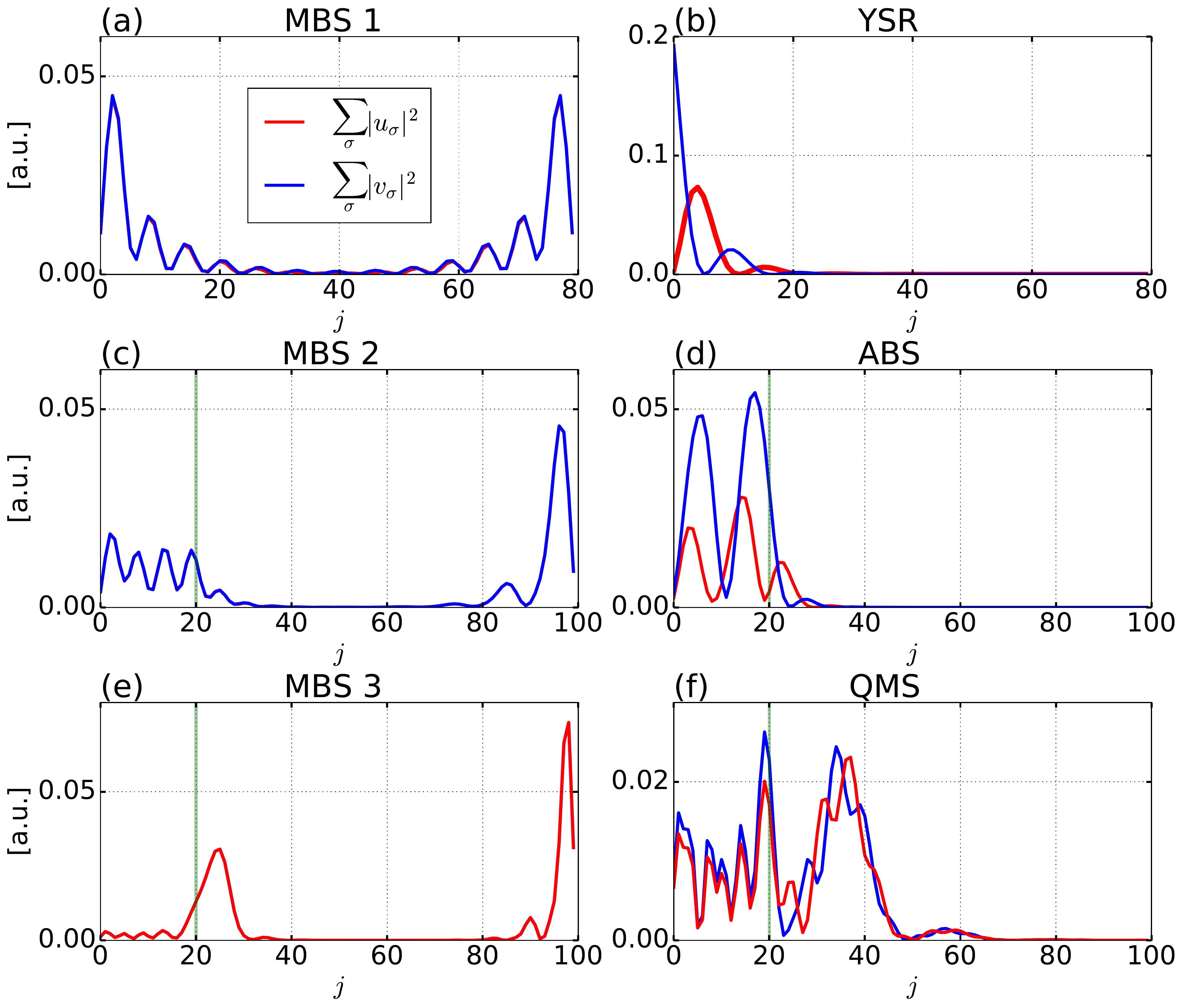}
    \caption{Local particle weight $\sum_\sigma |u_\sigma|^2$ (red solid line) and the local hole weight $\sum_\sigma |v_\sigma|^2$ (blue solid line) of the 6 different zero-energy states of Table~1.}
    \label{Particle/hole_weights}
\end{figure}

To be complete, we have plotted in Fig~\ref{Particle/hole_weights} the local particle weight $\sum_\sigma |u_\sigma|^2$ (red solid line) and the local hole weight $\sum_\sigma |v_\sigma|^2$ (blue solid line) of the 6 different zero-energy states of Table~1. In the case of perfectly unpaired MBS ({\it c},{\it e}), the curves perfectly overlap leading to a zero local particle-hole asymmetry at each position. In theother case of overlapped MBS or trivial bound-states, the curves do not perfectly overlap everywhere resulting in a finite local charge detected as Fano factor oscillations above 1.
\bibliography{Biblio_Noise2}